\documentclass[twocolumn,secnumarabic,amssymb, nobibnotes, aps, prl, superscriptaddress]{revtex4}

\usepackage[utf8x]{inputenc} 
\usepackage{graphicx}
\usepackage{dcolumn}
\usepackage{bm}
\usepackage{amsmath}
\usepackage{amsfonts}
\usepackage{threeparttable}
\usepackage{color} 
\usepackage{epstopdf}
\usepackage[normalem]{ulem}

\begin{document}

\title{Determination of the Fermi surface and field-induced quasi-particle tunneling around the Dirac nodal-loop in ZrSiS}

\author{C.~S.~A. M\"{u}ller}

\affiliation{High Field Magnet Laboratory (HFML-EMFL), Radboud University, Toernooiveld 7, Nijmegen 6525 ED, Netherlands.}
\affiliation{Radboud University, Institute for Molecules and Materials, Nijmegen 6525 AJ, Netherlands.}

\author{T.~Khouri}
\affiliation{High Field Magnet Laboratory (HFML-EMFL), Radboud University, Toernooiveld 7, Nijmegen 6525 ED, Netherlands.}
\affiliation{Radboud University, Institute for Molecules and Materials, Nijmegen 6525 AJ, Netherlands.}

\author{M.~R.~van~Delft}
\affiliation{High Field Magnet Laboratory (HFML-EMFL), Radboud University, Toernooiveld 7, Nijmegen 6525 ED, Netherlands.}
\affiliation{Radboud University, Institute for Molecules and Materials, Nijmegen 6525 AJ, Netherlands.}

\author{S.~Pezzini}
\altaffiliation{Present address: Center for Nanotechnology Innovation, NEST, Istituto Italiano di Tecnologia, Piazza S. Silvestro 12, 56127 Pisa, Italy.}
\affiliation{High Field Magnet Laboratory (HFML-EMFL), Radboud University, Toernooiveld 7, Nijmegen 6525 ED, Netherlands.}
\affiliation{Radboud University, Institute for Molecules and Materials, Nijmegen 6525 AJ, Netherlands.}

\author{Y.-T.~Hsu}
\affiliation{High Field Magnet Laboratory (HFML-EMFL), Radboud University, Toernooiveld 7, Nijmegen 6525 ED, Netherlands.}
\affiliation{Radboud University, Institute for Molecules and Materials, Nijmegen 6525 AJ, Netherlands.}

\author{J.~Ayres}
\affiliation{H. H. Wills Physics Laboratory, University of Bristol, Tyndall Avenue, Bristol BS8 1TL, UK.}

\author{M.~Breitkreiz}
\affiliation{Dahlem Center for Complex Quantum Systems and Fachbereich Physik, Freie Universit\" at Berlin, 14195 Berlin, Germany}

\author{L.~M.~Schoop}
\affiliation{Department of Chemistry, Princeton University, Princeton, New Jersey 08544, USA}

\author{A.~Carrington}
\affiliation{H. H. Wills Physics Laboratory, University of Bristol, Tyndall Avenue, Bristol BS8 1TL, UK.}

\author{N.~E.~Hussey}
\affiliation{High Field Magnet Laboratory (HFML-EMFL), Radboud University, Toernooiveld 7, Nijmegen 6525 ED, Netherlands.}
\affiliation{Radboud University, Institute for Molecules and Materials, Nijmegen 6525 AJ, Netherlands.}
\affiliation{H. H. Wills Physics Laboratory, University of Bristol, Tyndall Avenue, Bristol BS8 1TL, UK.}

\author{S.~Wiedmann}
\email{steffen.wiedmann@ru.nl}
\affiliation{High Field Magnet Laboratory (HFML-EMFL), Radboud University, Toernooiveld 7, Nijmegen 6525 ED, Netherlands.}
\affiliation{Radboud University, Institute for Molecules and Materials, Nijmegen 6525 AJ, Netherlands.}\

\date{\today}

\begin{abstract}
Unambiguous and complete determination of the Fermi surface is a primary step in understanding
the electronic properties of topical metals and semi-metals, but only in a relatively few cases has this goal
been realized. In this work, we present a systematic high-field quantum oscillation study up to
35 T on ZrSiS, a textbook example of a nodal-line semimetal with only linearly dispersive bands crossing 
the Fermi energy. The topology of the Fermi surface is determined with unprecedented precision and all
pockets are identified by comparing the measured angle dependence of the quantum oscillations to density 
functional theory calculations. Comparison of the Shubnikov-de Haas and de Haas-van Alphen oscillations 
at low temperatures and analysis of the respective Dingle plots reveal the presence of significantly 
enhanced scattering on the electron pocket. Above a threshold field that is aligned along the $c$-axis of the 
crystal, the specific cage-like Fermi surface of ZrSiS allows for electron-hole tunneling to occur across finite 
gaps in momentum space leading to quantum oscillations with a complex frequency spectrum. Additional 
high-frequency quantum oscillations signify magnetic breakdown orbits that encircle the entire Dirac 
nodal loop. We suggest that the persistence of quantum oscillations in the resistivity to high temperatures 
is caused by Stark interference between orbits of nearly equal masses.

\end{abstract}


\maketitle

\section{Introduction}

The electronic properties of almost all metals and semi-metals of current interest, be they correlated, topological or both, 
are determined by their low-energy excitations, i.e. those in the vicinity of the Fermi level. Hence, precise knowledge of 
the underlying Fermi surface is a fundamental step in elucidating the physics of the material in question. Examples of 
such determinations in materials of current intrest, however, are few and far between. Within the sphere of correlated electron systems, a recent angle-resolved photoemission spectroscopy (ARPES) study of the layered perovskite Sr$_2$RuO$_4$ enabled the most precise estimate to date of the (correlation-modified) spin-orbit coupling \cite{Tamai2019}. A decade earlier, a combined study of angle-dependent 
magnetoresistance (ADMR) and quantum oscillations (QOs) revealed the entire three-dimensional structure of the Fermi surface 
of the overdoped cuprate Tl$_2$Ba$_2$CuO$_{6+\delta}$ \cite{Hussey2003,Rourke2010}. Earlier still, pioneering 
QO studies revealed a near-complete picture of the multiple Fermi surfaces of the unconventional heavy fermion superconductor 
UPt$_3$ \cite{Joynt2002}.

Topological materials are, for the most part, uncorrelated and as such, band structure calculations based on density functional 
theory (DFT) ought to provide an accurate estimate of the topological features of the Fermi surface such as protected linear band crossings. Confirmation of this association, however, is often hindered by the presence of additional carriers in the bulk in topological insulators or other bands with a quadratic dispersion relation in the band structure of topological semimetals. Bi$_2$Se$_3$ is a classic example. Although Bi$_2$Se$_3$ is arguably the simplest member of the family of three-dimensional topological insulators \cite{Chen2009}, 
accessing the topological surface states in transport is invariably obscured by a large residual carrier density in the bulk.
And while QOs are a powerful means to distinguish between bulk and surface charge carriers via their angle dependence \cite{Wiedmann2016}, 
their analysis and interpretation remain controversial. In the case of Weyl semimetals, there has been a great deal of 
spectroscopic evidence supporting the existence of Fermi arcs \cite{Jia2016}. However, candidate materials often display a 
complex band structure that contains multiple Weyl points as well as other topologically trivial Fermi pockets, thereby 
rendering incomplete our understanding of their electronic properties \cite{Armitage2018,Gao2019}.

In this contribution, we report on the full determination of the Fermi surface of the Dirac nodal-line semi-metal (NLSM) 
ZrSiS. NLSMs are not only ideal systems with which to investigate the properties of charge carriers with a linear 
dispersion relation, they also have the potential to create a platform for research on topological correlated matter. 
Theoretical predictions show that screening of the (long-range) Coulomb interaction is weaker in these 
materials compared to conventional metals owing to a vanishing density of states near the Fermi energy \cite{Huh2016}. 
This, in combination with the metallic nature of NLSMs, makes this material class more susceptible to the development of a macroscopically 
ordered state such as superconductivity, magnetism, charge density wave or even excitonic order \cite{Liu2017,Roy2017,Rudenko2018}. 
Initial indications of correlation effects in ZrSiS were reported in Ref.~\cite{Pezzini2018} while tip-induced superconductivity 
was reported recently \cite{Aggarwal2019}. ZrSiS also shows a number of other striking properties including an optical 
conductivity that is independent of frequency over a large frequency range \cite{Schilling2017} and a butterfly-shaped 
angle-dependent magnetoresistance \cite{Ali2016}. In order to account for all of its remarkable material properties, however, 
complete knowledge of the Fermi surface topology of ZrSiS is essential.

Here, we report the determination of the full Fermi surface topology of ZrSiS through a detailed and high-resolution study of de Haas-van 
Alphen (dHvA) and Shubnikov-de Haas (SdH) oscillations in high magnetic fields up to 35~T. All six extremal orbits predicted from the Fermi surface calculated by DFT are identified experimentally.  
In addition, we observe a plethora of other frequencies arising from magnetic breakdown corresponding 
either to linear combinations of adjacent pockets or to large orbits enclosing the area of an entire nodal loop that, when taken together,  
markedly constrain the geometry and location of the individual (electron and hole) Fermi pockets. Employing the temperature dependence 
of the dHvA oscillations, we demonstrate that the cyclotron masses of the breakdown orbits are in good agreement with theoretical predictions 
for electron-hole tunneling. We also assign the individual high-frequency orbits that are formed across gaps in the Fermi surface 
enclosing different windings around vertices. 

This study confirms ZrSiS to be a textbook NLSM in the sense that the Fermi surface is composed entirely of bands with a Dirac-like dispersion 
that form individual nodal loops in a cage-like network of band crossing lines close to the Fermi level truncated only by 
weak spin-orbit interaction. Certain features of the experimentally-derived Fermi surface, however, are not captured by DFT, 
possibly due to correlation effects that are not taken into account in the original band structure calculations. Moreover, 
a comparison between the QO results obtained by torque (determined by the single-particle response) and transport (determined 
by the collective response) provides evidence for scattering hot-spots on one of the pockets. Finally, it is revealed 
that the QO extending up to 100~K in the transport data \cite{Pezzini2018} have a different form from those seen at lower 
temperatures. We speculate here that the specific cage-like Fermi surface of ZrSiS provides the necessary condition 
for a quantum interferometer to be realized.

\section{Experimental details and analysis}

Single crystals of ZrSiS were grown by placing elemental ratios of Zr, Si and S together with a small amount of I$_2$ in a 
carbon coated quartz tube. The tube was sealed under vacuum and heated to 1100$^{\circ}$C for a week. A temperature gradient 
of 100$^{\circ}$C was applied. The resulting crystals were wrapped in Zr foil and annealed under vacuum at 600$^{\circ}$C for 3
weeks. The structure and composition were verified with powder X-ray diffraction and EDX on individual crystals. 

We employed capacitive torque magnetometry to measure dHvA oscillations. The sample was glued on top of a copper-beryllium plate and the capacitance 
between two plates was measured with an analogue capacitance bridge with a frequency operating at 50~kHz. For the transport 
experiments, we evaporated Ti and Au on four lines across the samples and attached Au wires of 25 and 50 $\mu$m diameter  
directly to the Au strips with a small drop of silver conductive paste (Dupont 4929). The resistance data were acquired 
in four-probe configuration with a constant current excitation of 1~mA using standard lock-in acquisition. Torque 
magnetometry and transport experiments have been performed on a rotating stage and cooled in a $^3$He system with a base 
temperature of 0.3~K. The field measurements were carried out in a resistive Bitter magnet at the HFML, with a maximum 
field strength of 35~T.  All Fast Fourier transforms (FFTs) presented here are analyzed using the Hann window function 
with an appropriate zero-padding.

\section{Fermi surface}

\begin{figure}
	\includegraphics[width=\linewidth]{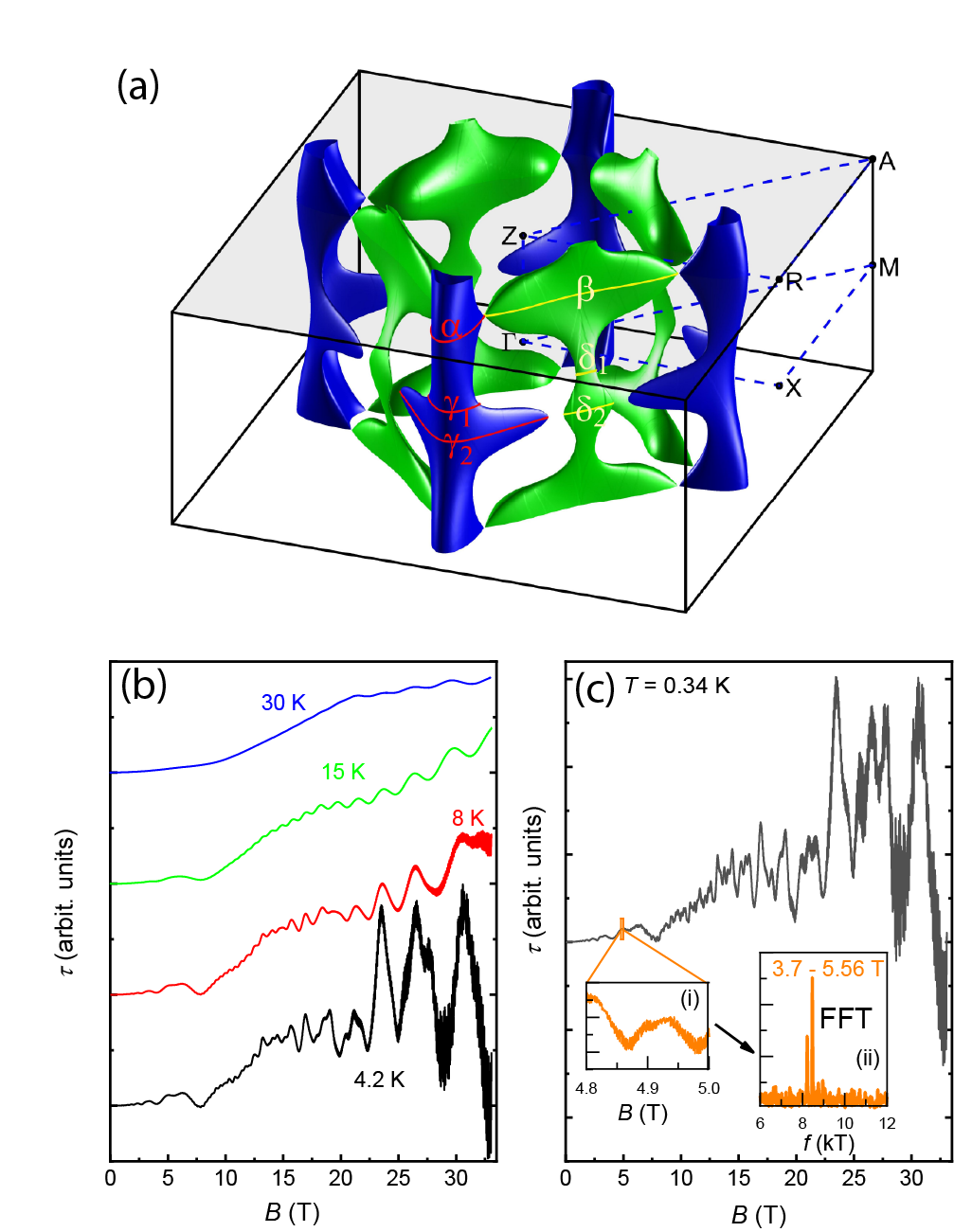}
	\caption{\label{Fig1} (a) 3D Fermi surface of ZrSiS. High symmetry points in the centre and the top surface of the Brillouin zone 
		are indicated. All pockets observed in the dHvA oscillations are labelled by greek letters. Electron and hole pockets are colored green and blue, respectively. (b) Torque $\tau$ as a function of 
		magnetic field for temperatures between 30 and 4.2~K and (c) at 0.34~K with $B$ oriented parallel to the $c$-axis of the crystal. 
		The signal is composed of quantum oscillations with multiple frequencies below 1~kT as well as a high frequency part. The insets 
		show a magnification of (i) $\tau$ in a regime between 4.8 and 5~T and (ii) a FFT performed between 3.7 and 5.56~T indicating 
		the presence of two frequencies above 8~kT.}
\end{figure}

ZrSiS is a member of the family of compounds with the general formula XSiY (X = Zr, Hf and Y = S, Se, Te). Its crystal 
structure (tetragonal space group P4/nmm) has a PbFCl structure type \cite{Schoop2016}. Layers of Zr and S are sandwiched 
between square nets of Si lying on glide planes (within the $ab$ plane) that provide the appropriate two-dimensional (2D) 
square lattice motif with non-symmorphic symmetry that Young and Kane first predicted would harbor topologically protected 2D Dirac 
cones \cite{Young2015}. In the absence of spin-orbit coupling (SOC), all bands near the Fermi energy $E_F$ in ZrSiS 
have a Dirac-like dispersion \cite{Schoop2016,Topp2017,Neupane2016,Chen2017} forming a line node in the Brillouin zone 
that gives rise to a cage-like Fermi surface. At the (high symmetry) X-point, there is a linear band crossing located around 0.7~eV below $E_F$
that is associated with the protected 2D Dirac fermions \cite{Schoop2016} and which survives the introduction of SOC. Other 
Dirac-like bands near $E_F$ \cite{Schoop2016} however become gapped in the presence of SOC. Surface states have also been
observed in ARPES experiments though these have been attributed to a reduced symmetry at the surface rather than to band 
inversion \cite{Topp2017}. Previous DFT calculations have generated many variations on the possible Fermi surface geometry of ZrSiS 
\cite{Pezzini2018,Ali2016,Chen2017,Fu2019,Lv2016,Novak2019,Su2018}. In Fig.~\ref{Fig1}(a), we show the three-dimensional (3D) 
Fermi surface plot of ZrSiS obtained from our own DFT calculations; for details see Supplemental Material \cite{SM}. As will be 
shown below, all elements of this DFT-derived Fermi surface are reproduced in our high-resolution dHvA study, thus 
giving us confidence that Fig.~\ref{Fig1}(a) captures the essential features of the Fermi surface that reproduces largely the
experimental results. (Note that while ARPES \cite{Schoop2016,Topp2017,Neupane2016,Chen2017} and FT-STS \cite{Lodge2017,Butler2017} have confirmed the diamond-like Fermi surface within the plane, it is essentially derived from the projected surface Brillouin zone - 
the so-called $\overline{\Gamma XM}$-plane). The Fermi surface of ZrSiS has a cage-like shape within the ($k_x$, $k_y$)-plane 
that is quasi-2D, i.e. open along $k_z$ but still strongly dispersive. For illustration, we label the individual electron 
and hole pockets which we identify in here in Fig.~\ref{Fig1}(a). In total, we identify six extremal orbits: $\alpha$ and $\beta$ in the \textit{Z-R-A}-plane, $\gamma_2$, $\delta_2$ in the \textit{$\Gamma$-X-M}-plane, while $\gamma_1$ and $\delta_1$ reside
between the \textit{Z-R-A} and \textit{$\Gamma$-X-M}-planes.

We now proceed with our experimental findings. In Fig.~\ref{Fig1}(b,c), torque $\tau$ as a function of the magnetic field $B$ 
aligned with the $c$-axis of the sample is shown for several chosen temperatures. With decreasing temperature, 
we observe a complex pattern of well-pronounced QOs. In the following, we present an analysis of the QO pattern that we 
divide into two parts. We will first focus on frequencies $f < $1~kT before turning our attention to high frequency QO that, 
in fact, develop already above 4.5~T at 0.34~K, as highlighted in inset (i) to Fig.~\ref{Fig1}(c). For illustration, the corresponding 
FFT in the field range between 3.7 and 5.56~T is shown in inset (ii) revealing two high frequencies above 8~kT.

\begin{figure*}
	\includegraphics[width=0.9\linewidth]{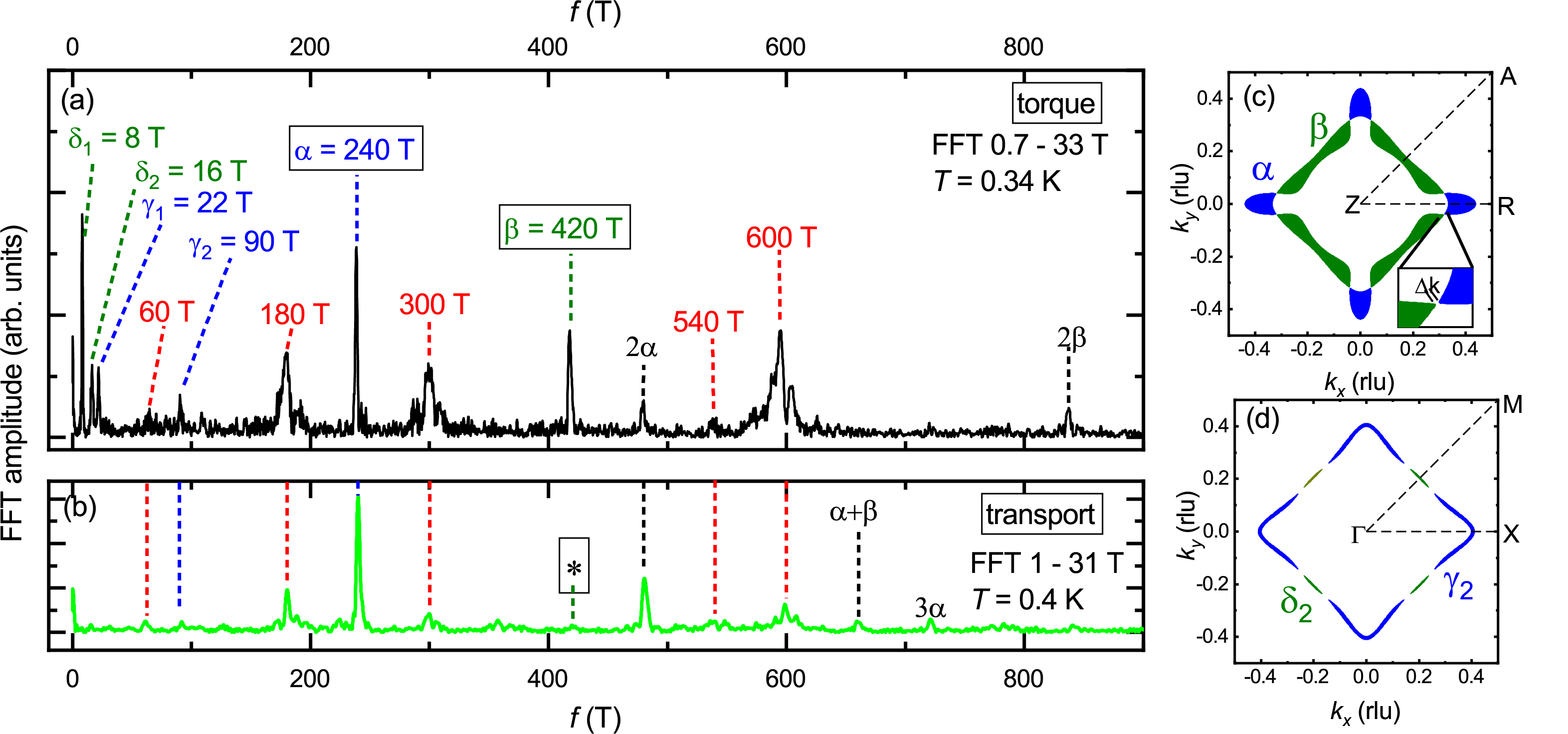}
	\caption{\label{Fig2} (a) Fast Fourier Transform of the dHvA oscillations at $T$= 0.34 K and (b) SdH oscillations at 0.4~K 
		in a frequency range below 1~kT. Fermi surface of ZrSiS in the \textit{Z-R-A}(c) and \textit{$\Gamma$-X-M} plane (d). Multiple frequencies 
		are visible that have been labeled and color coded according to their origin (electron or hole pockets). The $\alpha$ and $\beta$ frequencies 
		($f$ = 240 and 420~T respectively) originate from the two distinct electron and hole pockets in the \textit{Z-R-A}-plane as 
		shown in (c). These pockets are separated by a small gap in momentum space $\Delta k$ that arises due to spin-orbit coupling 
		(see inset).}
\end{figure*}

In total, we have measured both dHvA and SdH oscillations on three samples from the same growth. The resulting
FFTs ($f <$ 1~kT) obtained at the lowest temperatures are shown in Fig.~\ref{Fig2}(a) and (b) in a range from 0.7 to 33~T (31~T), respectively. Panels (c) and (d) also illustrate the projections of the DFT-derived Fermi surface at $\Theta $= 0$^{\circ}$ ($B \parallel c$-axis) within the \textit{Z-R-A} plane, 
and the \textit{$\Gamma$-X-M}plane. 

\begin{figure*}
	\includegraphics[width=0.9\linewidth]{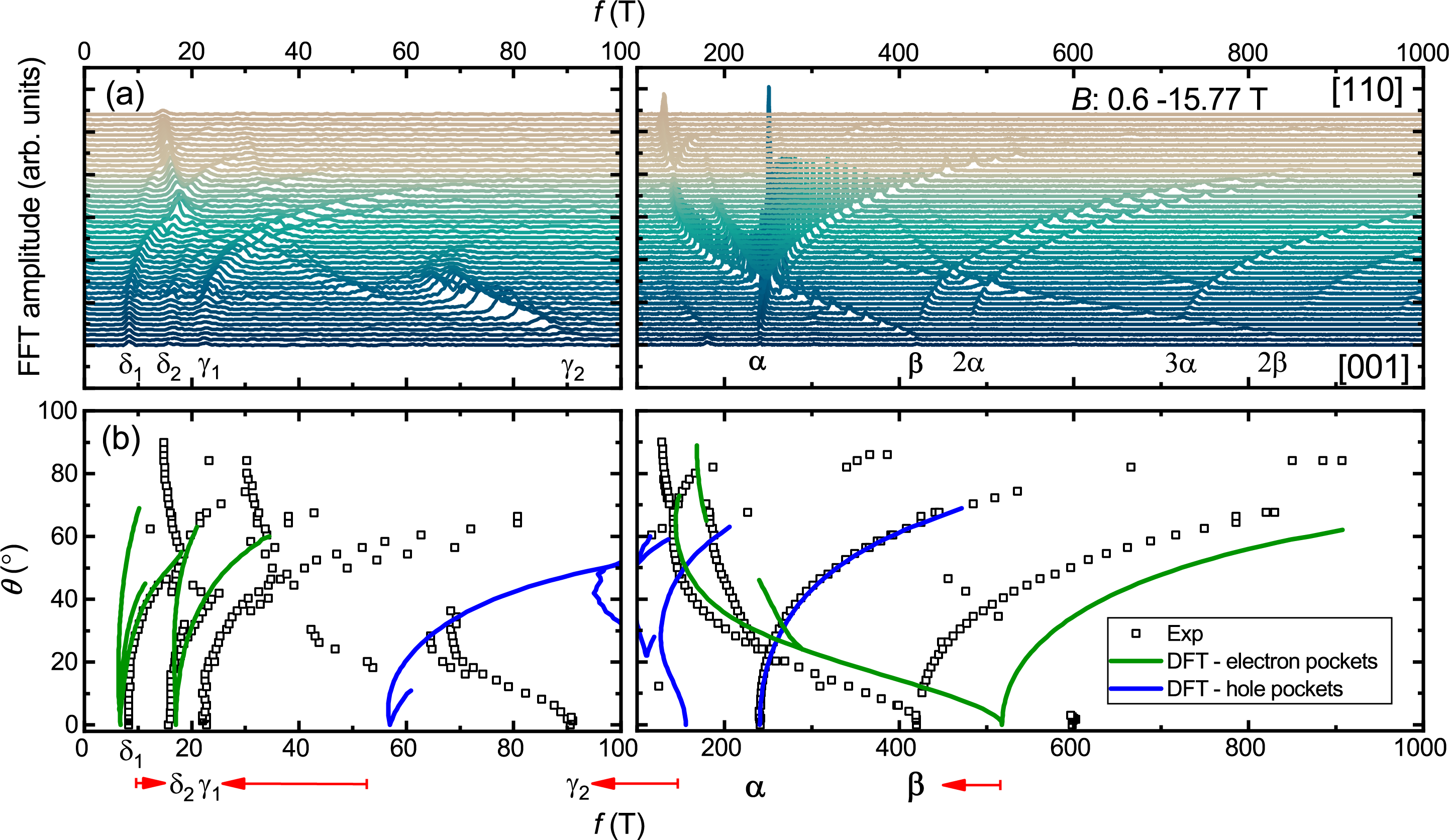}
	\caption{\label{Fig3} Angular dependence of the dHvA oscillations for $f < $1~kT: (a) Fast Fourier Transforms in the range from 0.6 to 15.77~T 
		taken from dHvA oscillations at $T$ = 1.2~K in steps of 2$^{\circ}$ from [001] to [110] - curves are offset for clarity. 
		(b) Comparison between the frequencies from the QOs (Exp.) and DFT calculations at different angles corresponding 
		to different elements (pockets) of the Fermi surface of ZrSiS.}
\end{figure*}

Multiple peaks are observed and have been labeled and color-coded according to their origin. Using the Onsager relation, 
we relate the frequencies $f$ observed in the FFT spectra to the extremal area $A_F$ of the individual pockets: 
$f = (\hbar/2\pi e) A_F$ \cite{Shoenberg1984}. All orbits with their corresponding frequency and mass are summarized 
in Table~\ref{tab1}. The cyclotron masses we determined by analyzing the temperature and field dependence of the QO signals, with an appropriate procedure adopted so that the finite field range does not affect the mass. Low-field measurements of the magnetization
of ZrSiS and the corresponding mass analysis including a general discussion on the QO analysis are presented in the SM, sections I and
IV, respectively \cite{SM}. The closed orbits of the Fermi surface are then deduced by comparing the experimental 
variation of the frequencies obtained from the angle dependence of the torque signal as $B$ is rotated from [001] to [110], 
illustrated in Fig.~\ref{Fig3}; for details see section II and the III in the Supplemental Material \cite{SM}. The comprehensive nature 
of our study allows us to unambiguously determine the individual pockets. Because our study covers such a large field range, 
we are able to observe all the fundamental frequencies present and therefore make a complete comparison with the DFT calculation.
For $f >$ 100~T, the largest amplitude peaks at 240~T and 420~T we label as $\alpha$ and $\beta$. We also observe 
their harmonics 2$\alpha$ and 2$\beta$ in torque, and a faint feature corresponding to 3$\alpha$ in transport. The $\alpha$-orbit 
is in excellent agreement with the theoretical prediction for the hole pocket located at the vertex of the Fermi surface 
in the \textit{Z-R-A }plane, see Fig.~\ref{Fig1}(a), and has been observed in all QO experiments to date 
\cite{Pezzini2018,Ali2016,Su2018,Singha2017,Matusiak2017,Hu2017,Zhang2018,Voerman2019}.

\begin{table}[t]
	\centering
	\caption{Identified orbits, extracted frequencies and extracted cyclotron masses from the experiment compared to the frequencies and cyclotron masses from the DFT calculations. The frequencies of the fundamentals are highlighted in bold text. Harmonics (2$\alpha$, 2$\beta$) have been omitted. The orbits are
		labeled in the corresponding planes of the Fermi surface in Figs.~\ref{Fig2}(c) and (d).}
	\label{tab1}
	\begin{tabular} {|c|c|c|c|c|}
		\hline
		Orbit				& $f$ (T) 			& $m_c~(m_e)$  			& $f_{DFT}$ (T)	& $m_{c;DFT}~(m_e)$  	\\
		\hline
		\hline
		$\delta _1$ 		&	\textbf{8}   	& $0.07\pm 0.01$ 		&	7 			& 0.07 			 		\\ 
		\hline
		$\delta _2$ 		&	\textbf{16}   	& $0.12\pm 0.01$ 		&	17 			& 0.11 					\\
		\hline
		$\gamma _1$ 		&   \textbf{22}  	& $0.19\pm 0.01$		&	57 			& 0.17					\\
		\hline
		2$\alpha -\beta$ 	&   60  			& -				 		&	- 			& - 				 	\\
		\hline
		$\gamma _2$ 		&   \textbf{90}  	& $0.30\pm 0.01$   		&	156 		& 0.33 				 	\\
		\hline
		$\alpha - \beta$ 	&  180 				& $0.73\pm 0.04$ 		&	- 			& - 					\\
		\hline
		$\alpha$ 			&  \textbf{240} 	& $0.19\pm 0.03$ 		&	240 		& 0.16 					\\
		\hline
		3$\alpha-\beta$   &  300 				& $1.02\pm 0.07$		&	- 			& - 				 	\\
		\hline
		$\beta$ 			& \textbf{420} 		& $0.62\pm 0.05$		&	518 		& 0.53 					\\
		\hline
		$4\alpha-\beta$   & 540 				&  - 					&	- 			& - 					\\
		\hline
		$2\beta -\alpha$    & 600 				& $1.22\pm 0.03$ 		&	- 			& - 					\\	
		\hline	
	\end{tabular}
\end{table}

Several other sharp peaks in the FFT spectrum are identified in Fig.~\ref{Fig2}(a) below 100~T and we associate two of these peaks 
with pockets within the \textit{$\Gamma$-X-M} plane, see Fig.~\ref{Fig2}(d): $\gamma_2$ at the corners of the plane and $\delta_2$, the small electron pocket in between.
The pockets $\gamma_1$ = 22~T and $\delta_1$ = 8~T correspond to extremal areas between the \textit{$\Gamma$-X-M} and \textit{Z-R-A} planes, see Fig.~\ref{Fig1}(a). Although $f(\delta_2) = 2f(\delta_1)$ at $\theta = 0^o$, this relation breaks down at finite angles. Hence, we can rule out the possibility that $f(\delta_2)$ is a harmonic of $f(\delta_1)$. We attribute $\delta_1$ to an electron-like orbit and $\gamma_1$ to a hole-like orbit, respectively, due to their angular dependence. 
Both frequencies are absent in the FFT spectrum deduced from the SdH oscillations, Fig.~\ref{Fig2}(b). We note that the frequencies for the hole pockets $\gamma_1$ and $\gamma_2$ are somewhat overestimated in the DFT calculation. Nevertheless, their angle dependence and cyclotron masses are in agreement with the experimental results. 

There have been many reports of QO in ZrSiS using a variety of different probes \cite{Pezzini2018,Ali2016,Lv2016,Su2018,Singha2017,Matusiak2017,Hu2017,Zhang2018,Voerman2019} that allows for a test of the 
validity of the DFT-derived band structure. Essentially, all of the previous studies were able to identify the $\alpha$ frequency
(240 $\pm$ 10~T) while certain studies could resolve some of the smaller frequencies, i.e. those below 30~T \cite{Lv2016}. However, none of the previous studies were able to resolve the main $\beta$ frequency (420 $\pm$ 10~T). From now on, we refer to the
$\alpha$- and $\beta$-pockets with frequencies $f_{\alpha}$ and $f_{\beta}$ as the `petal' and `dog-bone', respectively, as 
introduced in Ref. \cite{Pezzini2018} and illustrated in Fig.~\ref{Fig2}(c). 

It appears that the extremal orbit of the electron `dog-bone' pocket can only be identified unambiguously in the dHvA signal, 
the reason for which will be discussed below. In our DFT calculations, this orbit is predicted to give rise to a frequency of 
518~T at $\Theta $= 0$^{\circ}$. Hence, in our earlier SdH study, we had assigned an observed frequency at 600~T to the $\beta$-pocket, which was the only other peak in the FFT spectrum that could be assigned to a closed orbit \cite{Pezzini2018}. Similarly, in measurements of the thermoelectric power, a frequency of 583~T was also identified \cite{Matusiak2017}. The new assignment of the $\beta$-pocket is justified here 
by its dispersion with tilt angle (see Fig.~\ref{Fig3}), while the 600~T frequency is now believed to be related to a closed 
orbit due to tunneling of quasi-particles between adjacent $\alpha$ and $\beta$ pockets in the \textit{Z-R-A} plane (now identified 
as 2$\beta - \alpha$). It is noted that the size of the $\alpha$-pocket is perfectly captured by the DFT calculation, while the $\beta$-pocket is around 20 \% smaller. This may indicate that correlation effects (not captured in the DFT) are indeed larger along the dog-bone orbit which runs parallel to the nodal-line. Although the experimental frequency of the $\beta$-pocket is lower than the DFT predicted one at $\Theta$ = 0$^{\circ}$, the agreement improves considerably as $\Theta$ increases with angle and is in almost perfect agreement found for $\Theta\geq$~20$^{\circ}$, see Fig.~\ref{Fig3}(b). This corresponds to where this extremal orbit has moved away from the bulges (facing $Z$) in this pocket, see details in figures S2 and S3 in the SM (section II and III) \cite{SM}. Note that the orbit $\beta_2$ (which increases in frequency as $\Theta$ increases) is always higher than in the experiment as this orbit encircles the bulges at all angles. So apart from this slight difference in size of the bulges in this pocket the size and shape are well described by the DFT. 

The other peaks present in Figs.~\ref{Fig2}(a,b) correspond to magnetic breakdown orbits in the \textit{Z-R-A}-plane, which we discuss in the following section. Finally, two of the smallest frequencies (labeled $\gamma_1$ and $\delta_1$) are assigned to the minima in the $c$-axis dispersion of the electron and hole pockets that are localized in between the \textit{Z-R-A} and \textit{$\Gamma$-X-M} planes.

We now turn to address the marked differences in the FFT spectra displayed in Fig.~\ref{Fig2}(a) and Fig.~\ref{Fig2}(b) determined by dHvA and SdH respectively, in particular the absence of the peak corresponding to the $\beta$-pocket in the SdH-derived spectrum. We note here that no other SdH experiment has yet been able to identify a peak at this frequency. Its absence is made all the more puzzling by the appearance of peaks in the SdH FFT corresponding to breakdown orbits between $\alpha$ and $\beta$, e.g. at 180~T and 600~T, implying that SdH orbits involving quasi-particle tunnelling across multiple SOC induced gaps are more coherent than those that simply involve cyclotron motion around the dog bone. An analysis of the field dependence of the dHvA signal from the $\alpha$ and $\beta$-orbits (see SM \cite{SM} figure S4) reveals a striking difference in coherence of these orbits. For the $\alpha$-orbit there is no discernible (Dingle) impurity scattering within our noise level, whereas for the $\beta$-orbit, the scattering is at least one order of magnitude greater. Such a difference affects the width of the Landau levels and this affects the magnitude of the SdH and dHvA signals equally. Normally, it is assumed that magnitude of the SdH oscillations follow those in the density of states \cite{Adams1959}. If the scattering on the $\beta$-pockets was relatively isotropic around this orbit then it should be expected that its contribution to the total conductivity would be much smaller than for $\alpha$ and hence its oscillatory contribution would also be diminished. This would explain why the SdH signal from the $\beta$-pocket is absent. The observation of the breakdown orbits in the SdH signal could then be explained by contributions of the long-lived states on the $\alpha$-pocket, oscillating at the breakdown frequencies.  

Alternatively, the enhanced scattering on the $\beta$-orbit may result from very high scattering at a single point (hot spot). Such a point could be located close to the point where the $\alpha$ and $\beta$-orbits are closest, as at this point the Fermi velocity dips dramatically.  Such intense electron scattering at the vertex of a (reconstructed) Fermi pocket was also suggested to occur in underdoped cuprates \cite{Robinson2015} where charge order is believed to create pockets with strong curvature near the edge of their Fermi arcs \cite{Comin2014}.  If electrons tunnel across the breakdown gap they could avoid this scattering hot-spot, and hence this would explain why the breakdown orbits are observed but not the $\beta$-orbit itself. However, this alone would not explain the difference between SdH and dHvA, as such intense scattering at a single point would not affect significantly the total conductivity. However, exactly how such intense scattering would affect the SdH oscillation in a multiband system is not clear and further theoretical work on this point is required.
	
Finally, we note that in all previous SdH studies carried out on ZrSiS, the only other (fundamental) frequencies that were reported were 
those with frequencies less than 20~T and conceivably only those associated with the extremal areas of the warped, but continuous, Fermi cylinders located between the \textit{$\Gamma$-X-M} and \textit{Z-R-A} planes. Hence, a similar mechanism (enhanced scattering along the nodal line) might also be responsible for the other missing frequencies in the SdH results.  

\section{Magnetic breakdown between adjacent electron and hole pockets}

Having identified the fundamental frequencies of the FFT spectrum that correspond to elements of the Fermi surface, we now
consider the origin of the other peaks, color-coded in red in Fig.~\ref{Fig2}(a) that have not yet been assigned. 

As it turns out, all the other frequencies observed in Fig.~\ref{Fig2}(a) can be identified as linear combinations of $f_{\alpha}$ and $f_{\beta}$, which
we denote as $\left| n \alpha-p \beta \right|$ with $n$ and $p$ being integers. We note that these orbits are broadened in the FFT spectrum compared to the frequencies that correspond to the different individual pockets of the Fermi surface likely due the fact that they occur at higher magnetic fields and are thus generated by a smaller field window in the FFT. These specific orbits can thus be attributed to magnetic breakdown, specifically, quasi-particle tunneling between adjacent electron and hole pockets. The oscillation frequency of a breakdown orbit is given by the total enclosed
area, whereby the areas of the electron and hole pockets contribute with opposite signs. All assigned orbits are summarized in Table \ref{tab1}. The simplest semi-classically allowed breakdown orbit is $\beta-\alpha$ ($f$ = 180~T) that leads to QOs with 
a frequency set by the difference between $f_{\beta}$ and $f_{\alpha}$. A similar orbit was recently observed in the sister compound HfSiS in SdH oscillations \cite{vanDelft2018} and has the shape of a 'figure-of-eight' \cite{OBrien2016}.

The other frequencies in the FFT spectrum are more complex and require either $n$ and/or $p$ to be greater than unity. We identify in 
total three well-pronounced orbits with different combinations of $\left| n \alpha-p \beta \right|$ in Fig.~\ref{Fig2}(a): $2\beta-\alpha$ ($f$ = 600~T, formerly assigned to the $\beta$-pocket, see Fig.~\ref{Fig5}(c)), $2\alpha-\beta$ ($f$ = 60~T), $3\alpha-\beta$ ($f$ = 300~T), Fig.~\ref{Fig5}(b), and a faint feature at 540~T that corresponds to an orbit $4\alpha-\beta$. While the orbits at 60, 180, 300 and 600~T can also be identified in the FFT spectrum of the SdH oscillations, we note that only a clear peak at 600~T has been observed in our previous experiment (Ref.~\cite{Pezzini2018}). 
The higher quality data in transport reported here is attributed to improved crystal quality and sample preparation. 

\begin{figure}
	\includegraphics[width=\linewidth]{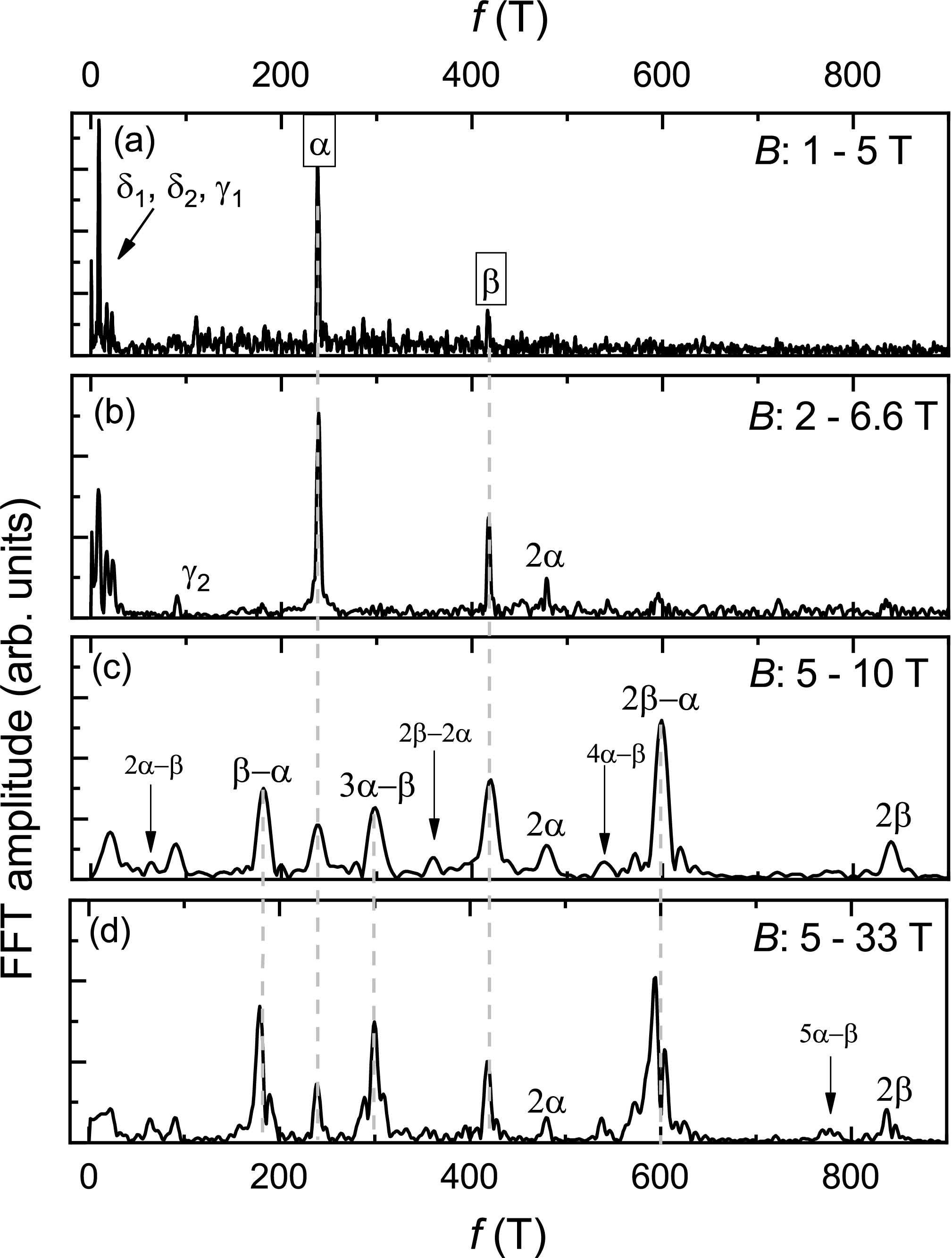}
	\caption{\label{Fig4} (a)-(d) Fast Fourier Transforms in different ranges of magnetic fields: in the low-field regime (a),
		only the frequencies that correspond to individual pockets ($\gamma_1, \gamma_2, \alpha$ and $\beta$) of the Fermi surface 
		are visible.}
\end{figure}

According to the DFT calculations, the electron and hole pockets in the \textit{Z-R-A} plane in ZrSiS are separated by a breakdown gap of around $\Delta k = 4.9 \times 10^{-3}$ \AA$^{-1}$ \cite{Pezzini2018}. This small gap permits tunneling to occur between adjacent pockets in sufficiently strong magnetic fields and gives rise a characteristic oscillation frequency given by the differences in frequencies associated with the individual pockets. The probability for tunneling to occur has an exponential dependence on the magnetic field, $\propto e^{-B_0/B}$, taking into account the number of breakdown events for a specific orbit. The breakdown field $B_0$ is $\propto (\Delta k)^2$ \cite{Shoenberg1984,Kaganov1983}. From DFT calculations, we find that the breakdown gap is in fact ten times larger in the \textit{$\Gamma$-X-M} plane. The reasons for this increase is two-fold. First the spin-orbit gap is larger by about a factor three in energy. 
The second reason is the Fermi velocity has a strong minimum at the point where the Fermi surfaces approach the point where the gap is smallest.
Given the exponential dependence on $\Delta k$, it ought to be too large for such electron-hole tunneling to be observed with the field strengths available to us in this study.

In Fig.~\ref{Fig4} we therefore analyse the dHvA oscillation spectrum, which illustrates the FFT amplitudes of the dHvA oscillation with frequencies below 1~kT shown in Fig.~\ref{Fig1}(c) over different magnetic field ranges. In the low-field range (1 to 5~T and 2 to 6.6~T), Fig.~\ref{Fig4}(a,b), only fundamental frequencies or their harmonics are clearly identified. By changing the range of the FFT analysis to higher fields, (5 to 10~T and 5 to 33~T), Fig.~\ref{Fig4}(c,d), we begin to uncover a progressively larger number of magnetic breakdown orbits labeled here with the corresponding linear combinations of $\alpha$ and $\beta$-pockets.

The second feature that is inextricably linked with magnetic breakdown due to electron-hole tunneling is the absolute value of the cyclotron
masses $m_c$ of the individual breakdown orbits. Following the theoretical prediction, the masses of combined electron and hole orbits add up as
\footnote{An extensive study of MB phenomena is presented in Ref.~\cite{Kaganov1983}. Taking into account that we deal with electron and hole carriers, i.e. $d(A)/dE$ and the ‘circulation’ direction of the carriers for closed orbits, we determine eq.~\ref{eq1}, see also Ref.~\cite{vanDelft2018}.}

\begin{equation}\label{eq1}
m_{n \alpha-p \beta} = n|m_\alpha|+p|m_\beta|.
\end{equation}

\begin{figure}
	\includegraphics[width=\linewidth]{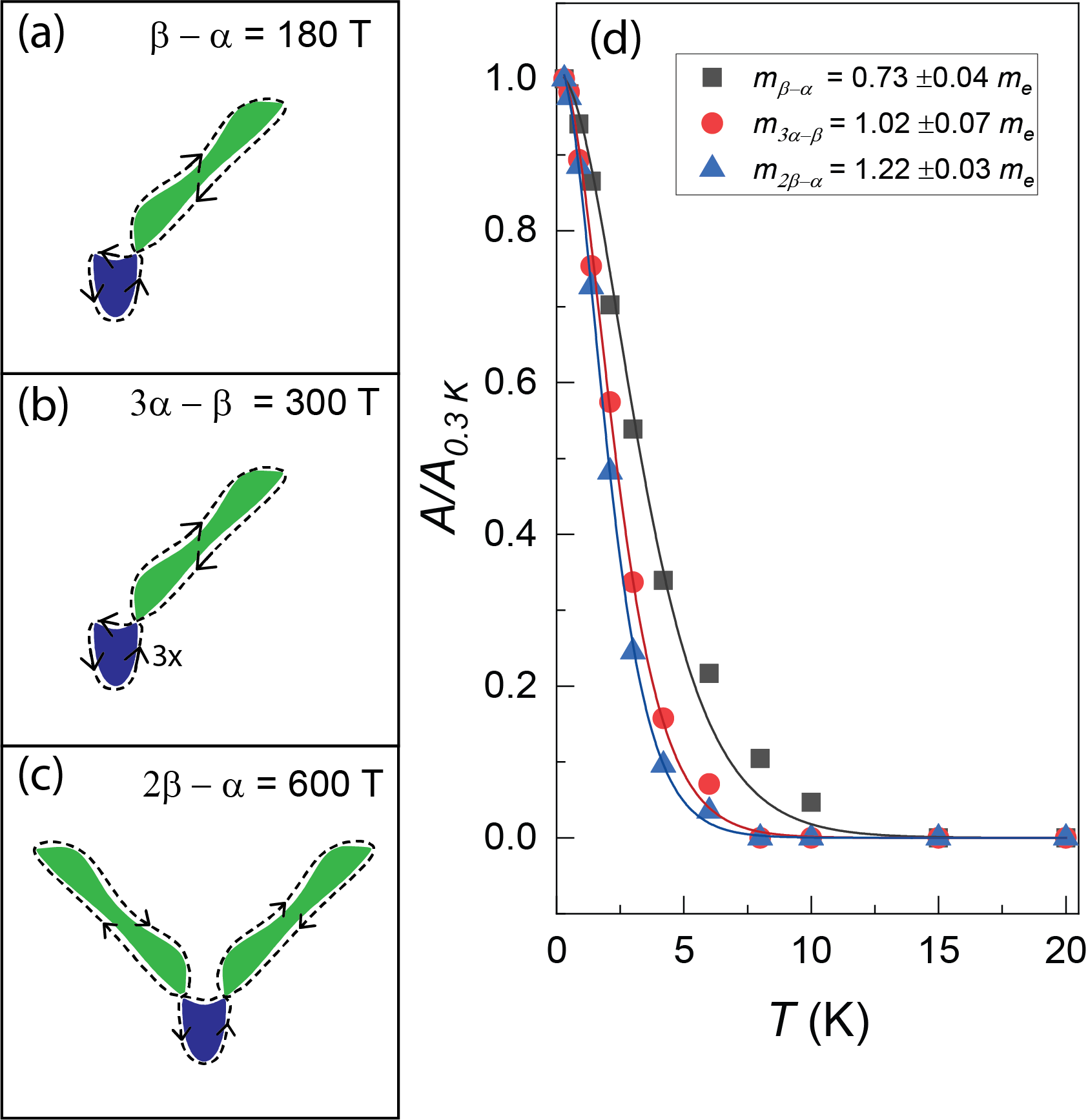}
	\caption{\label{Fig5} Sketches for three chosen breakdown orbits: (a) $\beta-\alpha$ (180~T), (b) $3\alpha-\beta$ (300~T) and
		(c) (c) $2\beta-\alpha$ (600~T). (d) Mass plot for the breakdown orbits (a)-(c). The resulting cyclotron masses $m_c$ are 
		specified in Table~\ref{tab1}.}
\end{figure} 

Figure~\ref{Fig5}(d) illustrates the evolution of the FFT amplitude normalized to the lowest temperature for three chosen
orbits $\beta-\alpha$, $3\alpha-\beta$ and $2\beta-\alpha$ corresponding to the combination of closed orbits depicted in Fig.~\ref{Fig5}(a-c).
Using the same procedure as described in Ref. \cite{vanDelft2018}, we deduce $m_c$. The results for all orbits are summarized 
in Table \ref{tab1}. Taking into account the uncertainty, we find that Eq. (\ref{eq1}) agrees well with the sum of
cyclotron masses from the individual pockets involved in the breakdown orbit. The third experimental signature for this phenomenon 
is the vanishing of all QOs associated with magnetic breakdown upon a slight tilting of the magnetic field with respect to the $c$-axis 
of the crystal. This has been demonstrated previously in both HfSiS and ZrSiS and supported by theoretical calculations \cite{Pezzini2018,vanDelft2018}.

\section{Comparison to theoretical expectations}

To compare the amplitudes of the signals (peaks in the FFT amplitude that correspond to magnetic breakdown orbits) with theoretical 
predictions, we follow Falicov and Stachowiak \cite{Falicov1966}, who considered a similar problem of a system of orbits coupled 
via magnetic breakdown applied in that case to Mg. Omitting a constant prefactor, the amplitude at a given frequency $f$ is given by 
\begin{align}
A ={}& B^\frac{5}{2}\sqrt{\sum_{ij} a_i a_j}, \label{ampl} \\
a_j ={}& \frac{w_j}{m_j^{5/2}}\big( ip \big)^{n_{pj}}\big( q\big)^{n_{qj}}e^{-\frac{ m_j}{\tau\omega_c }}
\frac{m_j\tfrac{T}{T_0}}{\sinh m_j\tfrac{T}{T_0}}, \label{ampl2}
\end{align}
where the sums in Eq. \eqref{ampl} run over all non-equivalent orbits of that frequency while the parameters are defined as follows:

\begin{center}
\begin{tabular}{ll}
   $p$ & tunneling amplitude absolute value $p=e^{-B_0/B}$ \\  
   $q$ &  same for avoided tunneling, $q=\sqrt{1-p^2}$ \\  
    $n_{pj}$ & number of tunnelings \\
 $n_{qj}$ & number of avoided tunnelings \\
  $T/T_0$ & temperature in units of $T_0 \approx 1 \mathrm{K} /16 B[\mathrm{T}]  $ \\
  $\tau$ & scattering lifetime\\
 $w_j$ & statistical weight factor \\
 $m_{j}$ & cyclotron mass in units of the free-electron mass.\\
\end{tabular}
\end{center}

\begin{table}[ht]
	\caption{Identified orbits, extracted frequencies, number of tunnelings and avoided tunnelings, statistical weight factor, mass ratios.}
	\label{tab2}
	\begin{tabular} {|c|c|c|c|c|c|}
		\hline
		Orbit & $f$ (T) & $n_{p\,j}$ & $n_{q\,j}$  & $w_j/m_j$ & $m_j/m_e$ \\
		\hline
		\hline
		$\alpha$ 			& 240  & $0$  & $2$ & $1$ & $0.19$ \\
		\hline
		$\beta$  			& 420  & $0$  & $2$ & $1$ & $0.62$\\
		\hline
		$\beta -\alpha$ 	& 180  & $2$  & $2$ & $2$ & $0.73$ \\
	     \hline
	     $2\alpha - \beta$   & 60  & $4$  & $2$ & $1$ & $0.88$\\
						 	             &   & $2$  & $4$ & $2$ &  \\
	     \hline
		$3\alpha - \beta$ 	& 300  & $2$  & $6$ & $1$ & $1.02$\\
								 	             &   & $4$  & $4$ & $2$ & \\
		\hline
		$4\alpha - \beta$   & 540  & $2$ & $8$  & $1$ & $1.17$ \\
								 	          &   & $4$  & $6$ & $3$ &\\
	     \hline
		$2\beta -\alpha$ 	& 600   & $4$  & $2$ & $1$ & $1.22$ \\
						 	             &   & $2$  & $4$ & $2$ & \\
		\hline
	\end{tabular}
\end{table}

The factor $i^{n_{pj}}$ in Eq. \eqref{ampl2} accounts for a phase shift of $\pi/2$ for each tunneling event \cite{Kaganov1983}. In contrast to the tunneling phase shift, Maslov phase shifts and topological phase shifts are always equal within the set of orbits at a given frequency and have therefore been omitted. 

\begin{figure}
	\includegraphics[width=\linewidth]{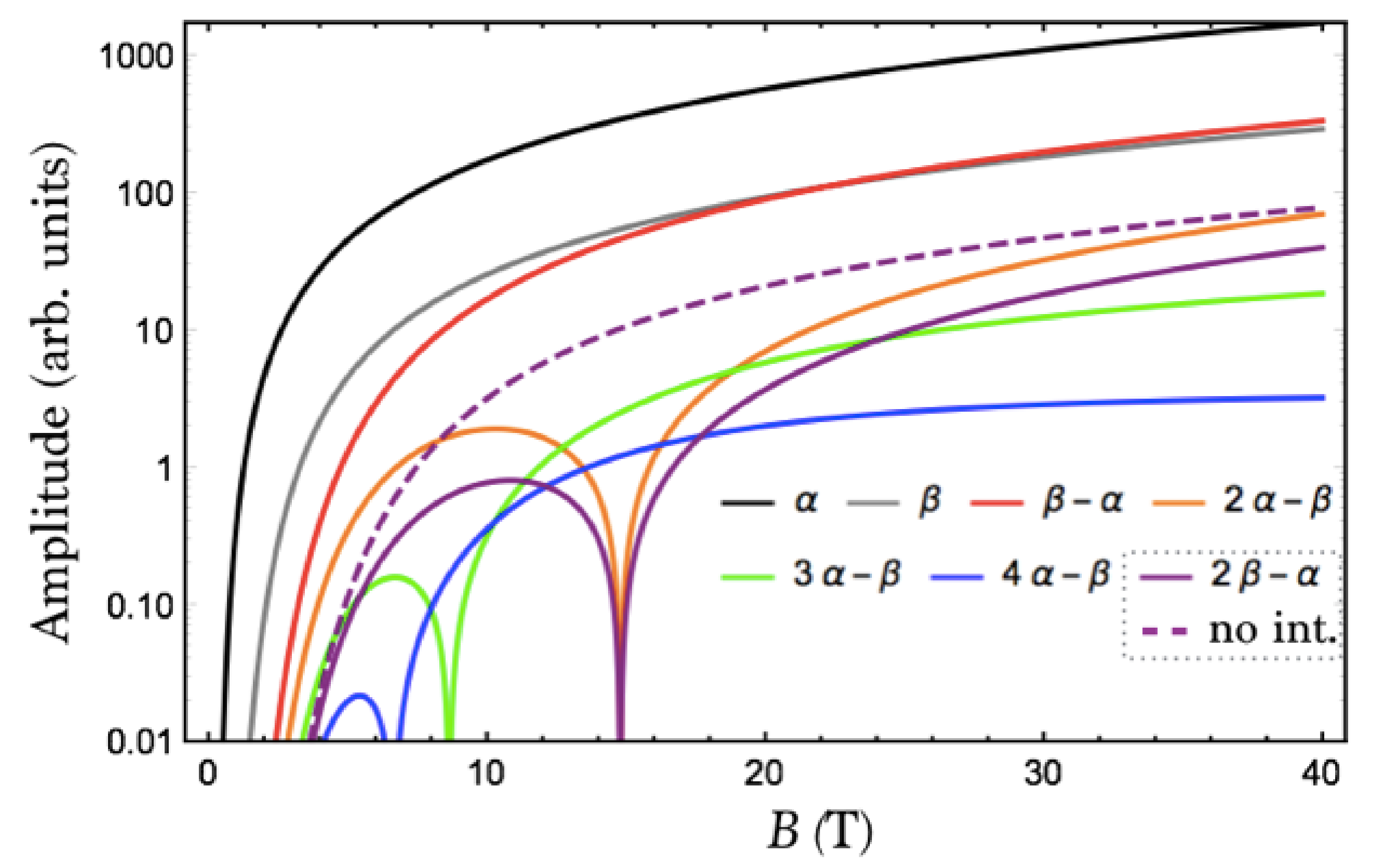}
	\caption{\label{Fig6} Theoretical $B$-dependence of the QO amplitudes for $B_0= 3$ T, $T=1$ K and $\omega_c\tau \gg 1$.}
\end{figure}

Figure \ref{Fig6} shows the $B$-dependence of the largest amplitudes for our system of coupled orbits with parameter values listed in Table \ref{tab2}. At small fields, the amplitudes are ordered as one would usually expect, with larger amplitudes for orbits with smaller cyclotron mass. At large fields, however, the amplitude of the heavy orbit $2\beta-\alpha$ becomes larger than that of the lighter orbits $3\alpha-\beta$ and $4\alpha-\beta$. The reason for this is that at large fields, tunneling is likely to occur and the amplitude becomes exponentially suppressed with the number of avoided tunnelings. The orbits  $3\alpha-\beta$ and $4\alpha-\beta$ involve at least $4$ and $6$ avoided tunnelings, respectively, while the orbit $2\beta-\alpha$ only $2$. 
 
\begin{figure}
	\includegraphics[width=0.4\linewidth]{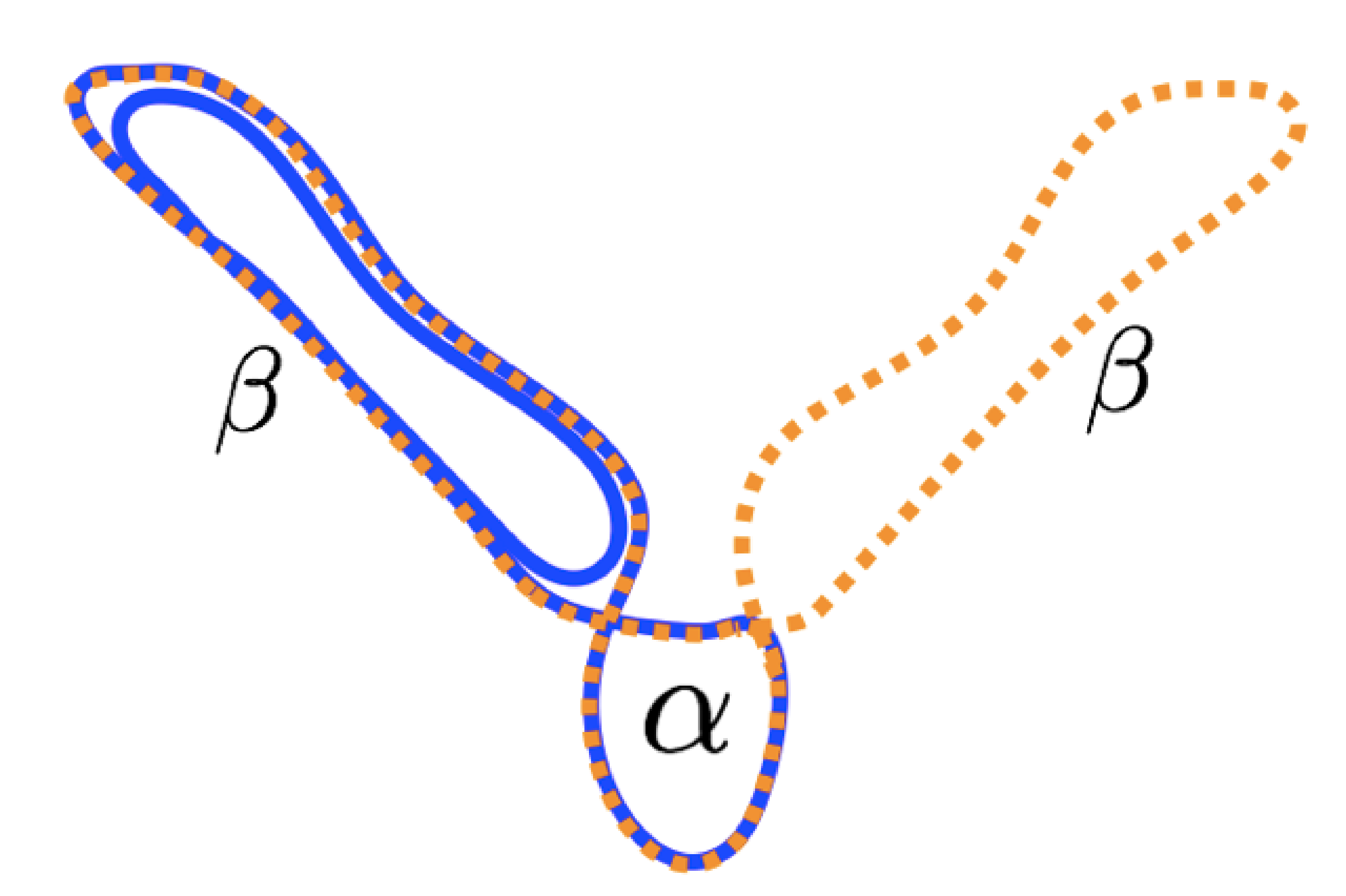}
	\caption{\label{Fig7} Interference of two orbits (dashed orange and solid blue) at the same frequency
	of $2\beta -\alpha$. The phase of the dashed orbit is expected to be shifted by $\pi$ due to the two extra
	 tunnelings, hence  leading to destructive interference with the other (solid) orbit. Similar considerations apply to all orbits with at least two 
	pockets of the same sort, such as $2\alpha -\beta$.}
\end{figure} 

Theory also predicts additional suppression at specific values of the $B$-field for orbits that involve at least two revolutions of an $\alpha$ or 
a $\beta$ pocket. The reason for this suppression is destructive interference of an orbit that evolves twice around the same pocket and an orbit at the same frequency that evolves around two different pockets of the same sort, as illustrated in Fig. \ref{Fig7}. These orbits have a phase difference of $\pi$ as the number of tunnelings differs by $2$. The comparison with the experimental spectrum in Fig.\ref{Fig4}(c) reveals a puzzling case however: while this interference effect can explain the strongly suppressed amplitude at the $2\alpha-\beta$ frequency, a similar suppression that is expected for $2\beta-\alpha$ is clearly absent. In fact, the amplitude of $2\alpha-\beta$  should dominate over $2\beta-\alpha$ at all fields due to the smaller mass, which is in contrast to our observations. This might indicate a so-far unknown deviation from the standard $\pi/2$ tunnel shift, possibly caused by the topological singularities in these materials \cite{Breitkreiz2018}. The dashed curve in Fig. \ref{Fig6} shows how the amplitude of $2\beta-\alpha$ would behave if the phase difference of the two orbits were zero, showing a much better agreement with the experiment.

\section{High-field magnetic breakdown}

\begin{figure}
	\includegraphics[width=\linewidth]{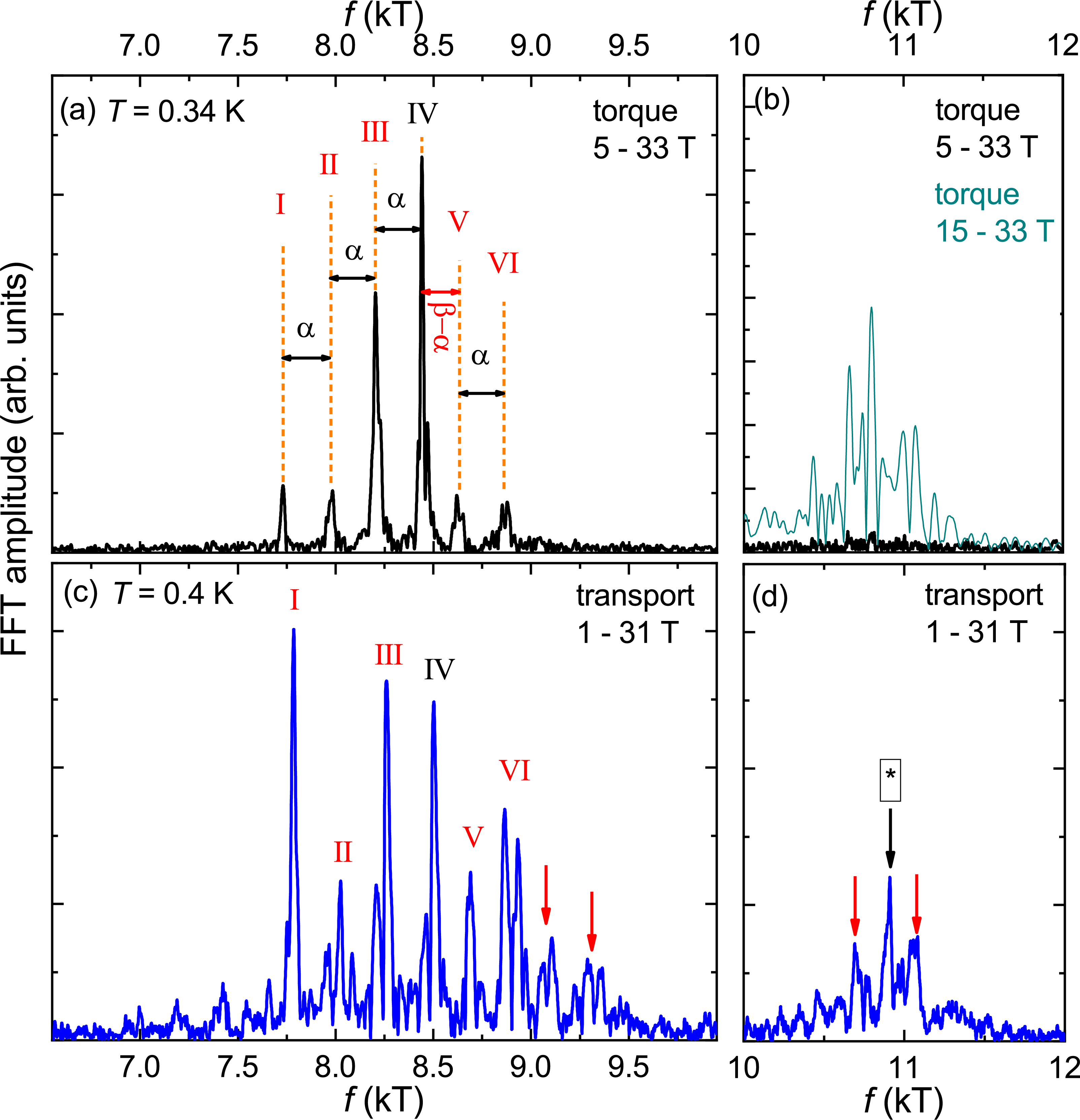}
	\caption{\label{Fig8} Fast Fourier Transforms for the high frequency oscillations in torque (a,b) and transport (c,d)
		at $T$ = 0.34 and 0.4~K, respectively. The spectrum is divided into two parts: a group around 8.6~kT and another group 
		around 11~kT. The high-frequency peaks (I)-(VI) in the 8.6~kT region are separated by individual $\alpha$'s and one
		$\beta - \alpha$.}
\end{figure}

We now turn to discuss the high-frequency QOs in ZrSiS first reported in \cite{Pezzini2018}. In contrast to the previous SdH study, the
oscillations are visible already for $B < $5~T as shown in inset
(ii) to Fig.~\ref{Fig1}(c) at 0.3~K. Figure~\ref{Fig8} shows the corresponding spectra of the FFT obtained from the dHvA (panels (a,b)) and SdH (panels (c,d)) studies. Both spectra reveal a series of peaks ranging from 7.5 to around 11.5~kT that can be divided into two groups: 7.5 to 9.5~kT and 10.5 to 11.5~kT.

In the FFT spectrum of the dHvA oscillations, Fig.~\ref{Fig8}(a), the most pronounced peaks appear between 8 and 9~kT and
are clearly separated from each other, in contrast to what is reported in Fig.~\ref{Fig8}(c) for the SdH-derived spectrum as 
well as in the previous SdH study \cite{Pezzini2018}. In total, we label six distinct peaks with Roman numbers I to 
VI with increasing frequency in this range. The separation in frequency between the peaks is indicated and corresponds 
to $\simeq$ 240~T, i.e. the frequency of the $\alpha$-pocket for all peaks except the one between peak V and IV which 
appears to correspond to the orbit $\beta - \alpha$. In view of these distinct separations, we can attribute the peaks in 
the FFT spectrum to magnetic breakdown orbits that reside in the \textit{Z-R-A} plane, see Fig.~\ref{Fig2}(b). 
The FFT spectrum from the SdH data also reveals peaks at the same 
frequencies though their relative amplitudes are different and they appear to be broader. Moreover, the overall resolution is 
reduced in comparison to the FFT spectrum of the dHvA data. In addition to the frequencies labeled from I to VI, 
two faint maxima (labeled by arrows) can also be discerned on the high-frequency side of this group. 

The second group between 10.5 to 11.5~kT have a much reduced amplitude. Within the resolution of the
torque and transport data, we are not able to determine a distinct separation of the FFT maxima by a certain frequency. The most
pronounced peak is found at 10.9~kT, marked with an asterisk in Fig.~\ref{Fig8}(d). In the previous SdH experiments
\cite{Pezzini2018}, we assigned this second group of peaks also to magnetic breakdown in the \textit{Z-R-A} plane, however, with the
precise determination of the Fermi surface and the full assignment of all pockets reported here, it is more likely that these higher frequencies are associated with loop orbits residing within the \textit{$\Gamma-X-M$} plane, which according to the DFT calculations, have a larger loop area, as shown in Fig.~\ref{Fig2}(d). According to the same DFT calculations, however, the gaps between the adjacent pockets $\gamma_2$ and $\delta_2$ are ten times larger than those in the \textit{Z-R-A} plane. Nevertheless, in the experiment, weak signatures of magnetic breakdown in the \textit{$\Gamma$-X-M} plane are clearly observed.

\begin{figure}
	\includegraphics[width=\linewidth]{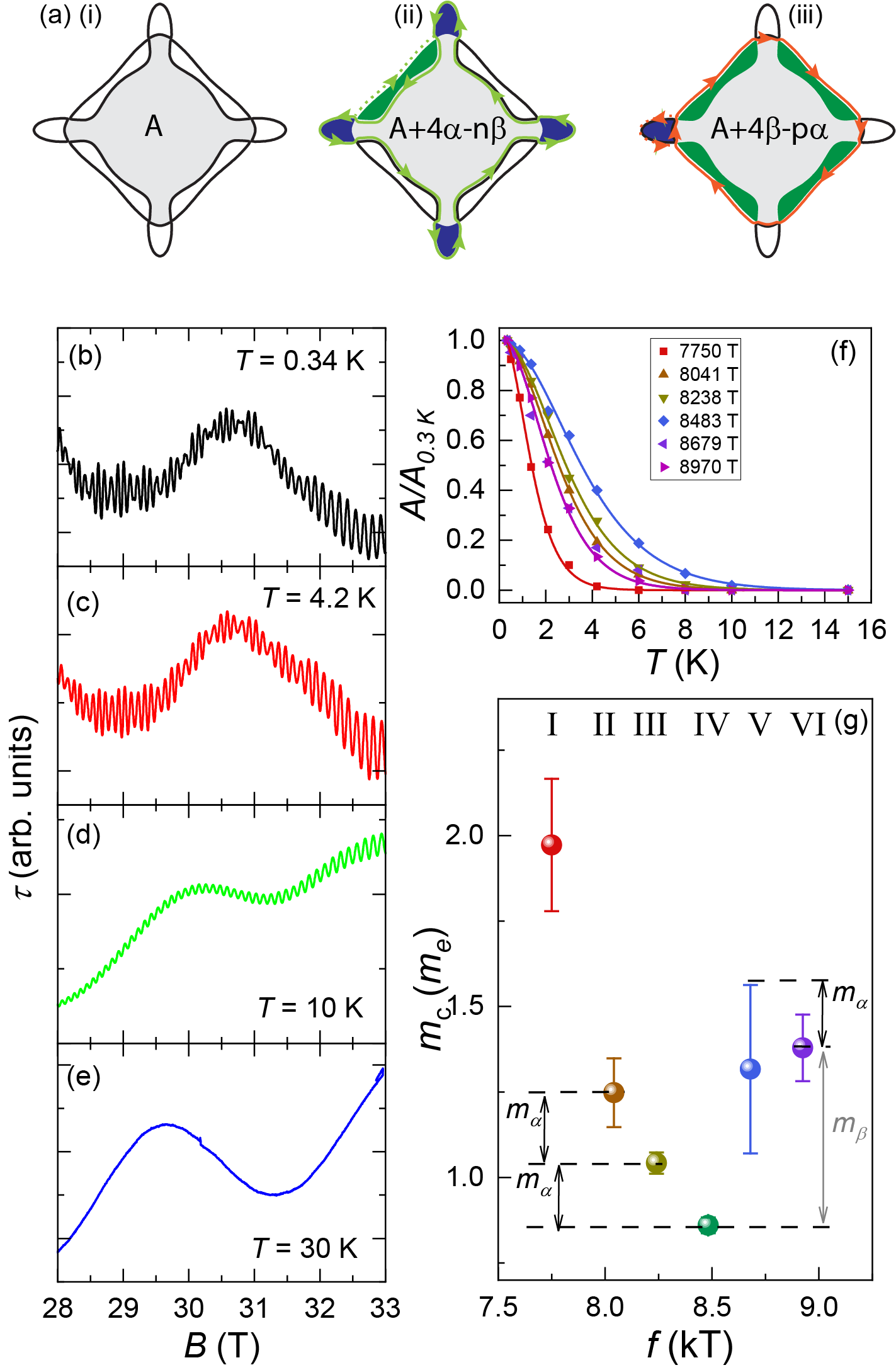}
	\caption{\label{Fig9} (a) Magnetic breakdown around the Dirac nodal loop within the \textit{Z-R-A} plane: sketch of different orbits including the inner area \textit{A}, and combinations of $\alpha$ and $\beta$-pockets. (b)-(e) Torque $\tau$ as a function of $B$ for several chosen temperatures. (f)
	Mass plot for all breakdown orbits of the 8.6~kT group. (g) Cyclotron masses $m_c$ for the breakdown orbits.}
\end{figure} 

Given the resolution of the high frequency oscillations in the \textit{$Z-R-A$ }plane, it is possible to identify the individual orbits responsible. A general description of the orbits associated
with the high frequency oscillations is shown in Fig.~\ref{Fig9}(a). In order to orientate the reader, we refer to the inner orbit in this plane as area 
\textit{A} (grey filled area in Fig.~\ref{Fig9}(a, i)). The distinct peaks labeled in Fig.~\ref{Fig8}(a) can now be defined 
by either including all the $\alpha$-pockets ($A + 4\alpha + p\alpha - n\beta$) or all the $\beta$-pockets ($A + 4\beta +n\beta - p\alpha$) with 
$n, p = 0,1,2,3, ...$ as illustrated in Figs.~\ref{Fig9}(a, ii and iii). Other combinations of orbits around $A$, such as $A + 2\alpha$ are semi-classically forbidden since they requires a reversal of the rotation direction of the quasi-particle. 

From temperature-dependent measurements of the dHvA oscillations, see Figs.~\ref{Fig9}(b-e), we extract the amplitude of the
corresponding individual frequencies, normalized to the lowest temperature, and plotted as a function of temperature in 
Fig.~\ref{Fig9}(f). A Lifshitz-Kosevich (LK) analysis for the thermal damping of QOs 
\cite{Shoenberg1984}, leads to the cyclotron masses $m_c$ for each of the observed orbits as shown in Fig.~\ref{Fig9}(g) 
and summarized in Table \ref{tab3}. The absolute values of $m_c$ are crucial for the assignment of the high-frequency
oscillations as we demonstrate below.

\begin{table}[ht]
	\caption{High-frequency orbits in the \textit{Z-R-A} plane, labeling, corresponding cyclotron masses, numbers of 
	tunnelings and avoided tunnelings, and statistical weight factor.}
	\label{tab3}
	\begin{tabular} {|c|c|c|c|c|c|c|}
		\hline
		peak 	& $f$ (T) & assignment  & $m_c\ (m_e)$ &$n_{p\, j}$ & $n_{q\, j}$& $w_ j/m_j$ \\
		\hline
		\hline
		I 		& 7730  & A + 4$\alpha$  						& $1.97\pm 0.19$ & $8$ & $0$ & $1$   \\
		&       & A + 4$\beta$ - 3$\alpha$  			& & $8$ & $6$ & $64$ \\
		\hline
		II  	& 7970  & A + 4$\beta$ - 2$\alpha$  			& $1.24\pm 0.10$  & $8$ & $4$ & $16$ \\
		\hline
		III 	& 8200  & A + 4$\beta$ - $\alpha$   			& $1.04\pm 0.03$ & $8$ & $2$ & $4$ \\
		\hline
		IV 		& 8440  & A + 4$\beta$  						& $0.86\pm 0.02$ & $8$ & $0$ & $1$\\
		\hline
		V 		& 8620  & A + 4$\beta$ + $\beta$ - $\alpha$   	& $1.31\pm 0.24$ & $8$ & $4$ & $16$ \\
		\hline
		VI   	& 8860  & A + 4$\beta$ + $\beta$ 			    & $1.38\pm 0.10$ & $8$ & $2$ & $4$ \\
		\hline
	\end{tabular}
\end{table}

In order to unravel the individual orbits associated with the high-frequency QOs, we start with peak IV ($f$ = 8440~T) which has 
the highest amplitude in the FFT spectrum of the dHvA oscillations and the lowest cyclotron mass. As will become clear below,
we define this orbit as $A + 4\beta$. Then, the lower frequencies with respect to $A + 4\beta$ correspond naturally to 
$A + 4\beta - n\alpha$ with $n$ =1 (peak III) and $n$ = 2 (peak II) since their frequencies are separated by 240~$\pm$~10~T 
and their corresponding $m_c$ increases by adding $n\cdot m_{\alpha}$ to the total cyclotron mass, respectively. Peak I ($f$ = 7730~T), in fact, is degenerate since the orbits $A + 4\alpha$ and 
$A + 4\beta - 3\alpha$ enclose the same area in momentum space. Following this analysis, it appears that the 
$A + 4\beta - 4\alpha$ is absent in the FFT spectrum of the dHvA oscillations.
On the high-frequency side, we find that the cyclotron masses of the orbits are also enhanced with respect to orbit IV. The assignment
of orbit VI with $m_{\rm IV} = A + 4\beta + \beta$ is justified by its observed cyclotron mass with $m_{A + 4\beta} + m_{\beta}$. The same holds for
orbit V ($A + 4\beta + \beta - \alpha$) with with $m_{\rm V}$ = $m_{A + 4\beta} + m_{\beta} + m_{\alpha}$. Taking these values, we can also
extract the experimental area for the inner orbit $A_{exp}$ = 6760~T. We note that the value is around 12 \% larger than
the area according to DFT estimate of 6~kT. A summary of the results is given in Table \ref{tab3}.

\begin{figure}
	\includegraphics[width=0.9\linewidth]{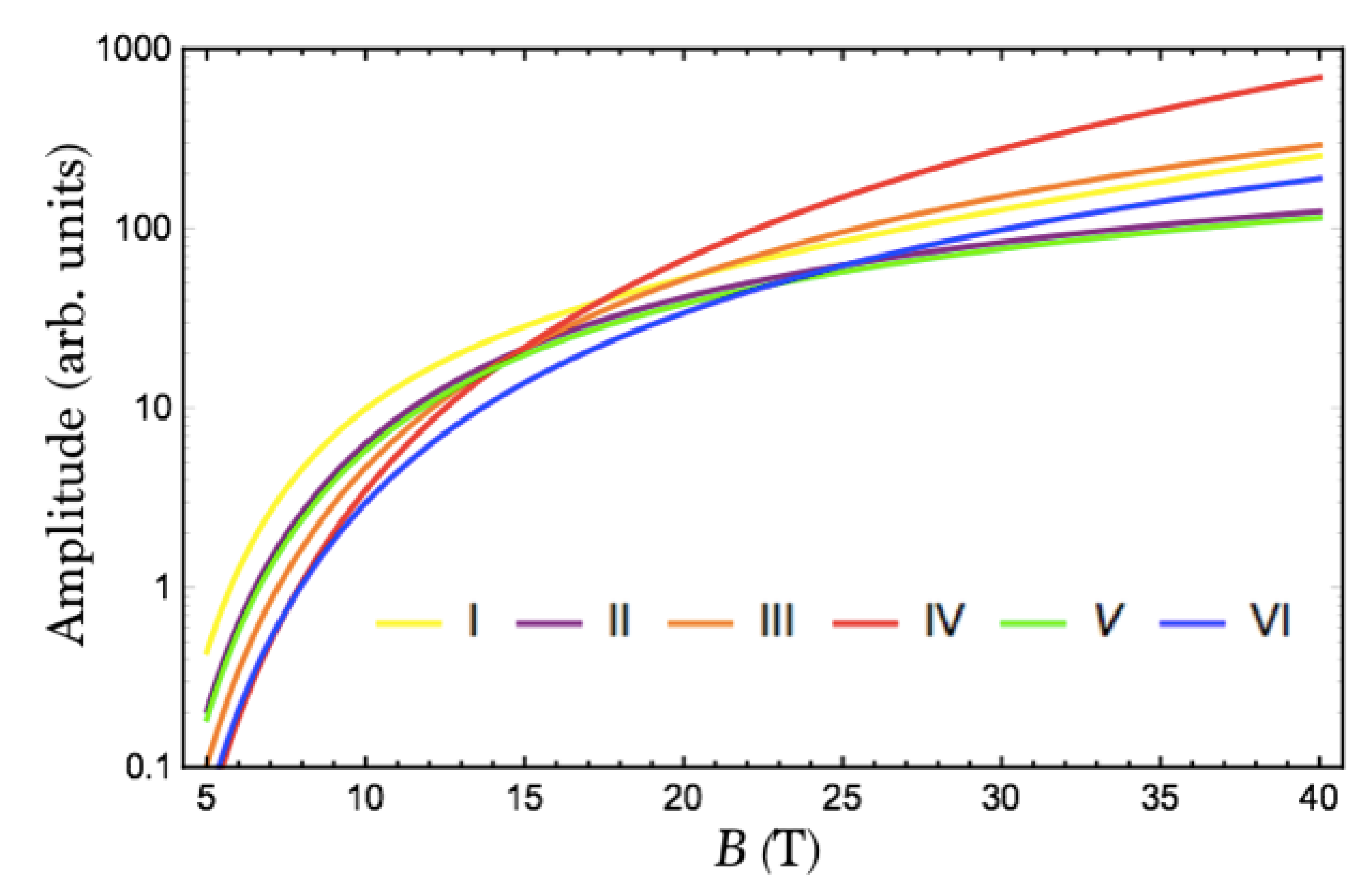}
	\caption{\label{Fig10} Theoretical $B$ dependence of the QO amplitude for $B_0= 3$ T, $T=0.3$ K and $\omega_c\tau \gg 1$.}
\end{figure} 

Figure \ref{Fig10} shows the calculated QO amplitudes with parameters given in Table \ref{tab3}. The amplitude differences are the result of the combined effect of the different masses, weight factors, and exponential suppression with the number of avoided tunnelings (while the number of tunnelings is equal for all the peaks). The order of the amplitudes, and especially the clear dominance of the frequency IV is in agreement with the experimental observation. 

\begin{figure}
	\includegraphics[width=0.9\linewidth]{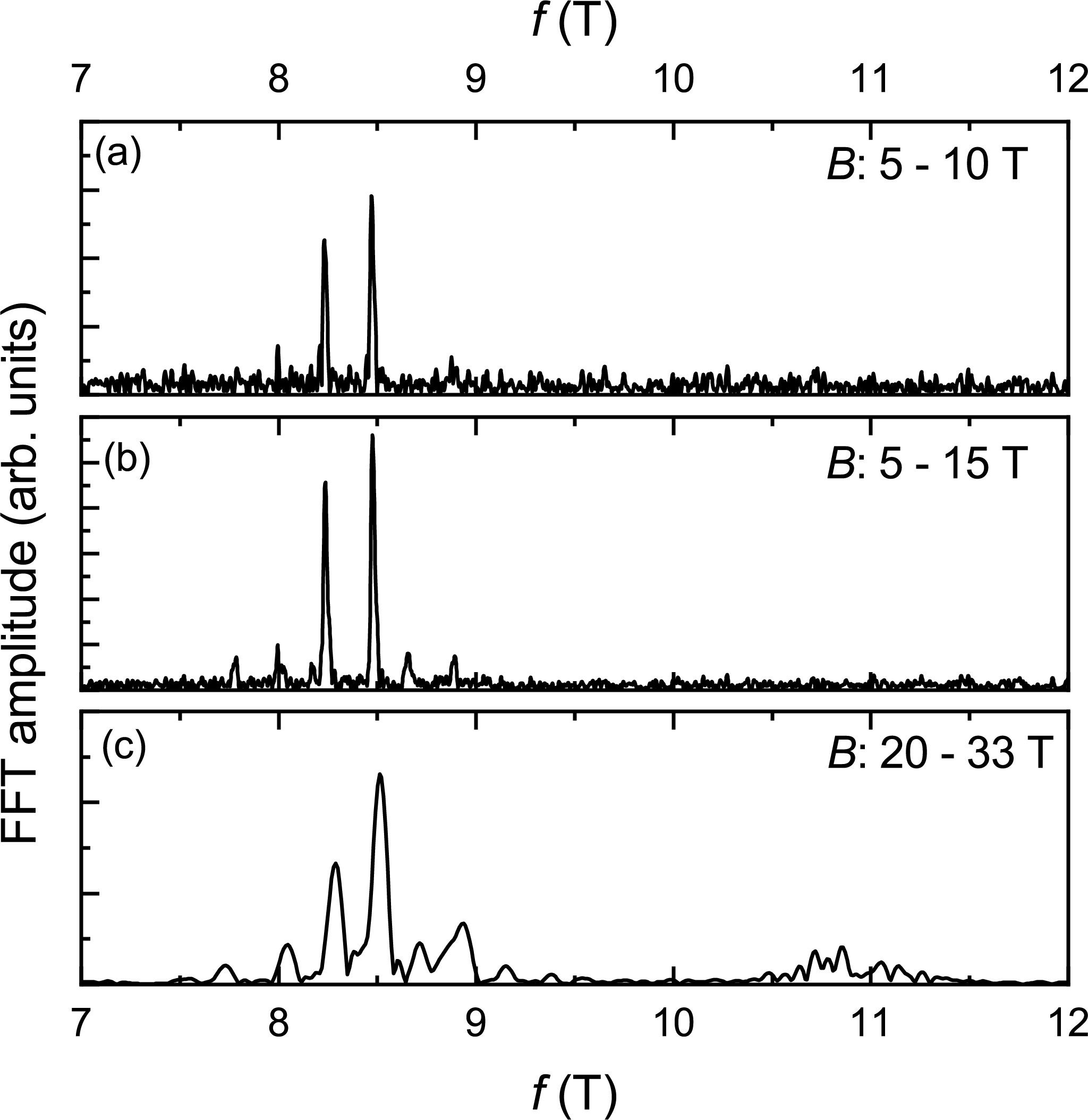}
	\caption{\label{Fig11} (a)-(c) Fast Fourier Transforms in different ranges of magnetic fields for the high-frequency magnetic breakdown orbits.}
\end{figure} 

Finally, we illustrate the onset field $B_0$ for the high-field QOs at $T$ = 0.34~K. Figure~\ref{Fig11} shows 
the FFT amplitudes of the dHvA oscillations in different magnetic field ranges between 7 and 12~kT. We note that the
QOs corresponding to the peaks in the FFT spectra do not appear in the same field-range, likely due to
an enhanced Dingle factor associated with the additional orbits that need to be traversed by the quasi-particles. Remarkably, 
the high-frequency oscillations appear already around 5~T, see Fig.~\ref{Fig1}(c).

\section{High temperature quantum oscillations}

\begin{figure}
	\includegraphics[width=0.9\linewidth]{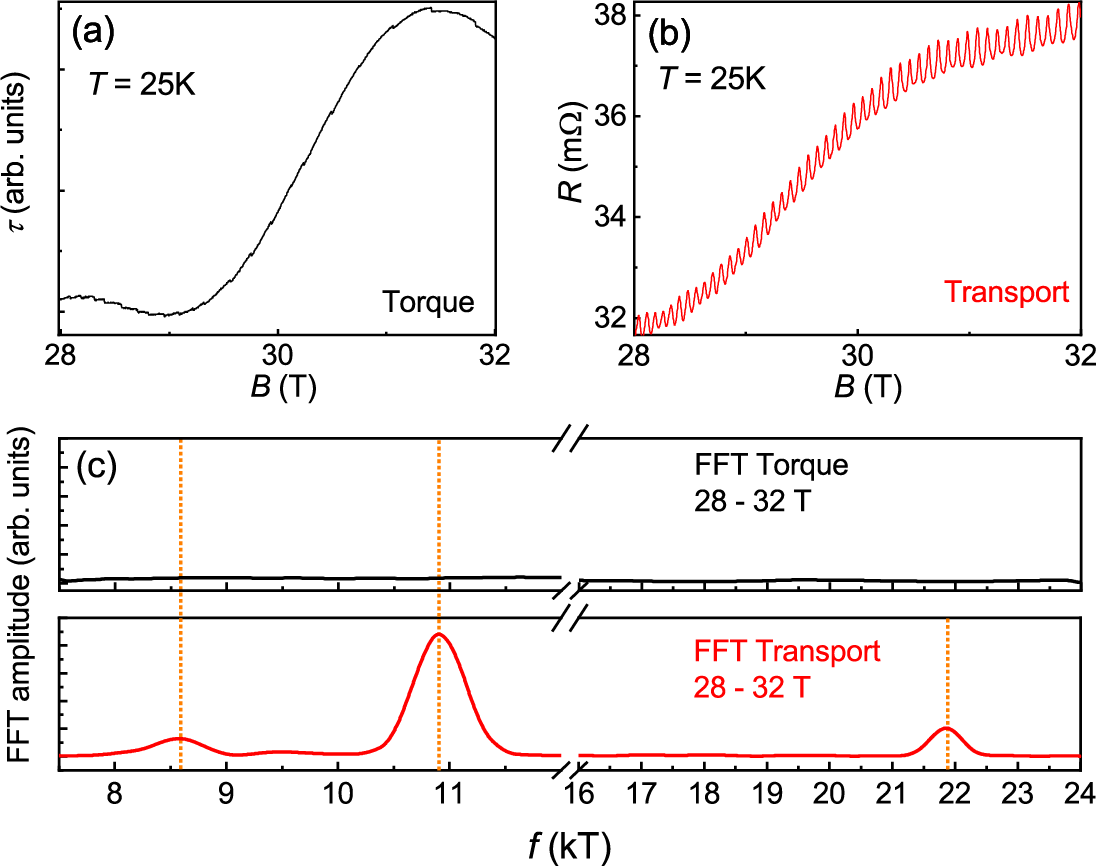}
	\caption{\label{Fig12} High-field torque signal $\tau$ (a) and transport (b) at 25~K. (c) Corresponding FFT spectra taken between 
	28 and 32 T. While the high frequency oscillations are absent in $\tau$, the frequencies at 8.6 and 11~kT (and its 1st harmonic) are
	identified in the FFT spectrum in transport.}
\end{figure}

As shown in Fig.~\ref{Fig9}(b-e), the high-frequency dHvA oscillations vanish with increasing temperature and follow the conventional LK
expression (Fig.~\ref{Fig9}(f)). This contrasts markedly with what was found in the previous SdH study \cite{Pezzini2018} where the high-frequency QOs were found to persist up to temperatures of order 100~K. Moreover, the $T$-dependence of their amplitude could be described by two LK expressions; one with a low mass to account for the high-$T$ oscillations and one with a high mass. In Figs.~\ref{Fig12}(a,b), we compare the dHvA and SdH QO data between 28 and 32~T at $T$~=~25~K, respectively. 
While the high-frequency dHvA oscillations are absent, we observe a pronounced oscillation pattern in transport. The corresponding 
FFT spectrum, Fig.~\ref{Fig12}(c), reveals in total three peaks at 8.6, $\sim$11 and $\sim$22~kT. The peak at 22~kT is the first harmonic of the 11~kT frequency. Looking back at the low-temperature FFT spectrum of the 
SdH oscillations in Fig.~\ref{Fig8}(c-d), we note that the orbits at 8.6 and 11~kT are present at low temperature though experience a small 
shift towards higher frequency.  Thus, we conclude that the orbits must necessarily originate from the \textit{Z-R-A} (8.6~kT) and \textit{$\Gamma$-X-M} (11~kT) planes, respectively.

\begin{figure}
	\includegraphics[width=\linewidth]{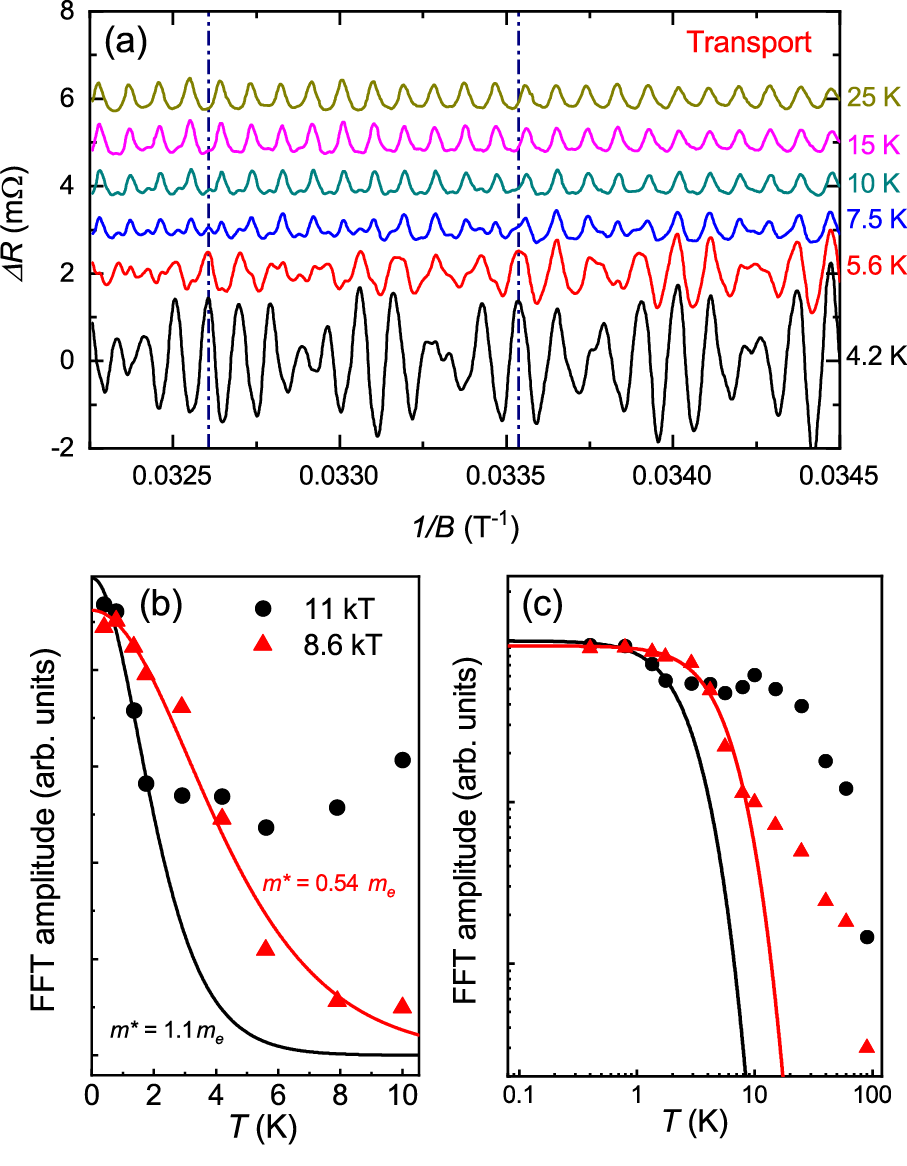}
	\caption{\label{Fig13} (a) High-field QOs in resistance $\Delta$R at different chosen temperatures from 4.2 to 25~K
		as a function of the inverse magnetic field 1/$B$. (b,c) FFT amplitude of the QOs as a function of temperature in different ranges.
		A clear deviation for the canonical Lifshitz-Kosevich expression (solid lines) is observed in the transport signal.}
\end{figure}

We now turn to consider the temperature dependence of the high-frequency oscillations in transport illustrated in Fig.~\ref{Fig13}(a). 
For this, we have removed the background originating from the low-frequency oscillation as well as the magneto-resistance at all temperatures 
and simply show oscillatory resistance $\Delta R$ as a function of 1/$B$ for several chosen temperatures. Several observations are made. 
Firstly, it is found that the shape of the oscillations changes markedly with increasing temperature. Secondly, while the oscillatory 
magneto-resistance clearly exhibits the shape of conventional SdH oscillations at 4.2~K, it appears that the oscillations
at 25~K exhibit sharp maxima and shallow minima. Thirdly, the periodicity and/or phase 
is changing with increasing temperature as indicated by the dash-dotted lines in Fig.~\ref{Fig13}(a).

As mentioned above, the temperature dependence of the SdH oscillations can be divided into two parts: a low-temperature regime shown 
in Fig.~\ref{Fig13}(b) ($T < 10$~K) and a high-temperature regime, Fig.~\ref{Fig13}(c), up to 100~K. The solid line corresponds 
to the canonical Lifshitz-Kosevich expression. Whilst we find a good agreement for the 8.6~kT frequency up to 10~K, the 
FFT amplitude of the 11~kT orbit strongly deviates from the canonical expression already above 2~K. At high temperatures, a clear
deviation is observed for both high-frequency orbits. This deviation has been attributed previously to exciton formation around an onset 
temperature of 8 K \cite{Rudenko2018}.

We note here that the overall shape of the high-$T$ QOs is reminiscent of the oscillatory resistance found in Mg and referred to there as a quantum 
interferometer, or Stark interferometer \cite{Stark1974,Stark1977}. Such QOs can only be observed in transport experiments, consistent 
with our experimental observations. The asymmetric shape of the QOs at high temperatures is due to the superposition of the dominant fundamental 
frequency (11~kT) and its first harmonic, which is well-pronounced in the FFT spectrum, see Fig.~\ref{Fig12}(c).

The general idea of a Stark interferometer is that the probability of a quasi-particle 
to traverse between two points in real or, equivalently, in momentum space can oscillate due to interference if there are two or 
more different paths connecting these points. For the interference effect to survive averaging over all points, the two path must differ by closed loops. To survive thermal broadening, the two paths must be of nearly equal length, in which case the energy variation of the phase difference cancels. 
Since the length of a closed orbit is proportional to the cyclotron mass, we conclude that Stark interference is possible if there are two (or more) different closed cyclotron orbits of nearly equal mass. Comparing the masses in Tables \ref{tab1} and \ref{tab3}, we identify three pairs of such orbits 
in which one is from the low and the other from the 8~kT high-frequency group listed in Table \ref{tab4}. 

\begin{table}[ht]
	\caption{Pairs of high- and low-frequency peaks with nearly equal masses. The mass of 
	$2\alpha-\beta$ could not be determined from the experiments and was therefore interpolated from the values of other orbits in Table \ref{tab1}. Note that the sign corresponds to the sign of the enclosed area, i.e., whether the orbit is electron- or hole-type. The frequency of Stark oscillations is given by the difference of the signed areas, since the magnetic-field induced phase changes are opposite for electrons and holes.}
	\label{tab4}
	\begin{tabular} {|c|c|c|c|c|c|}
		\hline
		high-$f$ peak 	& $f$ (T) 	& low-$f$ peak 			& $f$ (T) 	&   $\Delta m_c\ (m_e)$ & $\Delta f$ \\
		\hline
		\hline
		II  			& 7970  	&  $2\beta- \alpha$  	& $600$		& $0.02\pm 0.14$  		& $7370$ \\
		\hline
		III 			& 8200  	&  $3\alpha-\beta$		&  $-300$   &$0.02 \pm 0.10$ 		& $8500$  \\
		\hline
		IV 	 			& 8440  	& $2\alpha-\beta$     	& $-60$     &	 $0.02$ (interp.) 	& $8500$   \\
		\hline
	\end{tabular}
\end{table}

Our dHvA measurement shows that the probability of orbits II and $2\alpha-\beta$ is strongly suppressed, which suggests that only the pair III  ($3\alpha-\beta$) should be dominant. We conclude that the feature around $8500$ T in the high-temperature SdH data is Stark interference of two paths, one involving the III ($A+4\beta-\alpha$) and the other the $3\alpha-\beta$ orbit \footnote{The orbit with the 11~kT frequency and its first harmonic are the dominant contributions in the FFT, see Fig.~\ref{Fig12}(c), causing the fast oscillations at high temperatures. However, we do not observe pairs of high- and low-frequency oscillations and can therefore not identify possible combinations of orbits giving rise to the Stark interferometer in the \textit{$\Gamma$-X-M}-plane.}.

\section{Conclusions}

To conclude, we have presented a comprehensive QO study of the nodal-line semimetal ZrSiS. Notably, all six fundamental frequencies predicted by our DFT band-structure calculations were observed. While their dispersions with angle were captured extremely well by the calculations, some of the fundamental frequencies were found to be shifted relative to the DFT prediction. This shift implies the presence of some modified interaction term that is not captured fully in the current model. However, the observation of multiple breakdown orbits, some involving figure-of-eight orbits, others involving orbits around the entire Dirac nodal loop, constrains the resulting Fermi surface topology to such an extent that we are confident that the Fermi surface sketched in Fig.~\ref{Fig1}(a) represents the most accurate determination of the full Fermi surface of ZrSiS presently available. It also implies that many of the other proposals for its Fermi surface topology \cite{Chen2017,Fu2019,Lv2016,Novak2019,Su2018} are incorrect. Overall, the specific shape of the Fermi surface in ZrSiS is ideal to investigate the phenomenon of magnetic breakdown between adjacent electron and hole pockets. Furthermore, we have identified all orbits that lead to the high frequency dHvA oscillations in terms of their cyclotron masses. The distinct high-temperature quantum oscillations in transport and their anomalous temperature dependence are attributed to a Stark interferometer in the \textit{Z-R-A}-plane of the Fermi surface consisting of two different closed cyclotron orbits with nearly equal masses.

\begin{acknowledgments}
This work was supported by the High Field Magnet Laboratory - Radboud University/Foundation for Fundamental Research on Matter (HMFL-RU/FOM) a member of the European Magnetic Field Laboratory and by the UK Engineering and Physical Sciences Research Council (Grant No. EP/R011141/1). LMS has been supported by the Princeton Centre for Complex Materials, a Materials Science and Engineering Center (MRSEC) DMR 1420541. We thank Kamran Behnia for helpful discussions.
\end{acknowledgments}


\begin{thebibliography}{39}

\bibitem{Tamai2019}
A. Tamai, M. Zingl, E. Rozbicki, E. Cappelli, S. Ricco, A. de la Torre, S. McKeown Walker, F. Y. Bruno, P. D. C. King, W. Meevasana, M. Shi, M. Radovic, N. C. Plumb, A. S. Gibbs, A. P. Mackenzie, C. Berthod, H. U. R. Strand, M. Kim, A. Georges, and F. Baumberger, Physical Review X \textbf{9} 021048 (2019).

\bibitem{Hussey2003}
N. E. Hussey, M. Abdel-Jawad, A. Carrington, A. P. Mackenzie and L. Balicas, Nature \textbf{425}, 814 (2003).


\bibitem{Rourke2010}
P. M. C. Rourke, A. F. Bangura, T. M. Benseman, M Matusiak, J. R. Cooper, A. Carrington and N. E. Hussey, New Journal of Physics \textbf{12}, 105009 (2010).

\bibitem{Joynt2002}
see review article by R. Joynt and L. Taillefer, Reviews of Modern Physics \textbf{74}, 235 (2002) and references therein.

\bibitem{Chen2009}
Y. L. Chen, J. G. Analytis, J.-H. Chu, Z. K. Liu, S.-K. Mo, X. L. Qi, H. J. Zhang, D. H. Lu, X. Dai, Z. Fang, S. C. Zhang, I. R. Fisher, Z. Hussain and Z.-X. Shen, Science \textbf{325}, 178 (2009)

\bibitem{Wiedmann2016}
S. Wiedmann, A. Jost, B. Fauqu\'{e}, J. van Dijk, M. J. Meijer, T. Khouri, S. Pezzini, S. Grauer, S. Schreyeck, C. Br\"une, H. Buhmann, L. W. Molenkamp and N. E. Hussey, Physical Review B \textbf{94}, 081302(R) (2016).

\bibitem{Jia2016}
S. Jia, S.-Y. Xu and M. Zahid Hasan, Nature Materials \textbf{15}, 1140 (2016).

\bibitem{Armitage2018}
N. P. Armitage, E. J. Mele, and A. Vishwanath, Rev. Mod. Phys. \textbf{90}, 015001 (2018).

\bibitem{Gao2019}
H. Gao, J\"{o}rn W. F. Venderbos, Y. Kim, A. M. Rappe, Annual Review of Materials Research \textbf{49}, 153-183 (2019).

\bibitem{Huh2016}
Y. Huh, E.-G. Moon and Y.-B. Kim,  Phys. Rev. B \textbf{93}, 035138 (2016).

\bibitem{Liu2017}
J. Liu and L. Balents, Phys. Rev. B \textbf{95}, 075426 (2017).

\bibitem{Roy2017}
B. Roy, Phys. Rev. B \textbf{96}, 041113(R) (2017).

\bibitem{Rudenko2018}
A. N. Rudenko, E. A. Stepanov, A. I. Lichtenstein, and M. I. Katsnelson, Phys. Rev. Lett. \textbf{120}, 216401 (2018).

\bibitem{Pezzini2018}
S. Pezzini, M. R. van Delft, L. M. Schoop, B. V. Lotsch, A. Carrington, M. I. Katsnelson, N. E. Hussey and S. Wiedmann,
Nature Physics \textbf{14}, 178 (2018).

\bibitem{Aggarwal2019}
L. Aggarwal, C. K. Singh, M. Aslam, R. Singha, A. Pariari, S. Gayen, M. Kabir, P. Mandal and G. Sheet, J. Phys. Condens. Matt. \textbf{31}, 485707 (2019).

\bibitem{Schilling2017}
M. B. Schilling, L. M. Schoop, B. V. Lotsch, M. Dressel and A. V. Pronin, Phys. Rev. Lett. \textbf{119}, 187401 (2017).

\bibitem{Ali2016}
M. N. Ali, L. Schoop, C. Garg, J. M. Lippmann, E. Lara, B. Lotsch and S. Parkin, Sci. Adv. \textbf{2}e160174 (2016).

\bibitem{Schoop2016}
L. M. Schoop, M. N. Ali, C. Stra{\ss}er, A. Topp, A. Varykhalov, D. Marchenko, V. Duppel, S. S. P. Parkin, B. V. Lotsch and C. R. Ast, Nature Commun. \textbf{7}, 11696 (2016).

\bibitem{Young2015}
S. M. Young and C. L. Kane, Phys. Rev. Lett. \textbf{115}, 126803 (2015).

\bibitem{Topp2017}
A. Topp, \textit{et al.}, Phys. Rev. X \textbf{7} 041073 (2017).

\bibitem{Neupane2016}
M. Neupane, I. Belopolski, M. M. Hosen, D. S. Sanchez, R. Sankar, M. Szlawska, S-Y. Xu, K. Dimitri, N. Dhakal, P. Maldonado, P. M. Oppeneer, D. Kaczorowski, F. Chou, M. Z. Hasan and T. Durakiewicz, Phys. Rev. B \textbf{93}, 201104(R) (2016).

\bibitem{Chen2017}
C. Chen, X. Xu, J. Jiang, S.-C. Wu, Y. P. Qi, L. X. Yang, M. X. Wang, Y. Sun, N. B. M. Schr\"oter, H. F. Yang, L. M. Schoop, Y. Y. Lv, J. Zhou, Y. B. Chen, S. H. Yao, M. H. Lu, Y. F. Chen, C. Felser, B. H. Yan, Z. K. Liu and Y. L. Chen, Phys. Rev. B \textbf{95}, 125126 (2017). 

\bibitem{Fu2019}
B.-B. Fu, C.-J. Yi, T.-T. Zhang, M. Caputo, J.-Z. Ma, X. Gao, B. Q. Lv, L.-Y. Kong, Y.-B. Huang, P. Richard, M. Shi, V. N. Strocov, C. Fang, H.-M. Weng, Y.-G. Shi, T. Qian and H. Ding, Sci. Adv. \textbf{5}, eaau6549 (2019) .

\bibitem{Lv2016}
Y.-Y. Lv, B.-B. Zhang, X. Li, S.-H. Yao, Y. B. Chen, J. Zhou, S.-T. Zhang, M.-H. Lu and Y.-F. Chen, Appl. Phys. Lett. \textbf{108}, 144201 (2016) .

\bibitem{Novak2019}
M. Novak, S. N. Zhang, F. Orbanic, N. Biliskov, G. Eguchi, S. Paschen, A. Kimura, X. X. Wang, T. Osada, K. Uchida, M. Sato, Q. S. Wu, O. V. Yazyev and I. Kokanovic, Phys. Rev. B \textbf{100}, 085137 (2019).

\bibitem{Su2018} 
C.-C. Su, C.-S. Li, T.-C. Wang, \textit{et al.} New Journal of Physics, \textbf{20} 103025 (2018).

\bibitem{SM} 
In the Supplemental Material, we present a magnetization study, details on the band structure calculations and a Dingle analysis for individual pockets of the Fermi surface.

\bibitem{Lodge2017}
M. S. Lodge, G. Chang, C.-Y. Huang, B. Singh, J. Hellerstedt, M. T. Edmonds, D. Kaczorowski, Md. M. Hosen, M. Neupane, H. Lin, M. S. Fuhrer, B. Weber and M. Ishigami, Nano Lett. 17, 7213 (2017).
 
\bibitem{Butler2017}
C. J. Butler, Y.-M. Wu, C.-R. Hsing, Y. Tseng, R. Sankar, C.-M. Wei, F.-C. Chou and M.-T. Lin, Phys. Rev. B \textbf{96}, 195125 (2017).

\bibitem{Shoenberg1984}
D. Shoenberg, \textit{Magnetic oscillations in metals} (Cambridge University Press, Cambridge, 1984).

\bibitem{Singha2017} 
R. Singha, A. Pariari, B. Satpati and P. Mandal, PNAS, 5077-5082 \textbf{114}, 2468 (2017).

\bibitem{Matusiak2017}
M. Matusiak, J. R. Cooper, and D. Kaczorowski. Nature Communications, \textbf{8} 15219 (2017).

\bibitem{Hu2017}
J. Hu, Z. Tang, J. Liu, Y. Zhu, J. Wei and Z. Mao, Phys. Rev. B \textbf{96}, 045127 (2017).

\bibitem{Zhang2018}
J. Zhang \textit{et al.}, Front. Phys. \textbf{13}, 137201 (2018).

\bibitem{Voerman2019}
J. A. Voerman, L. Mulder, J. C. de Boer, Y. Huang, L. M. Schoop, C. Li and A. Brinkman, Phys. Rev. Mater. \textbf{3}, 084203 (2019). 

\bibitem{Adams1959}
E. N. Adams and T. D. Holstein, J. Phys. Chem. Solids \textbf{10}, 254 (1959).

\bibitem{Robinson2015}
P. Robinson and N. E. Hussey, Phys. Rev. B \textbf{92}, 220501(R) (2015).

\bibitem{Comin2014}
R. Comin, A. Frano, M. M. Yee, Y. Yoshida, H. Eisaki, E. Schierle, E. Weschke, R. Sutarto, F. He, A. Soumyanarayanan, Y. He, M. Le Tacon, I. S. Elfimov, J. E. Hoffman, G. A. Sawatzky, B. Keimer and A. Damascelli, Science \textbf{343}, 390 (2014).

\bibitem{vanDelft2018}
M. R. van Delft, S. Pezzini, T. Khouri, C. S. A. M\"{u}ller, M. Breitkreiz, L. M. Schoop, A. Carrington, N. E. Hussey, and S. Wiedmann,
Phys. Rev. Lett. \textbf{121}, 256602 (2018).

\bibitem{OBrien2016}
T. E. O'Brien, M. Diez, and C.W.J. Beenakker, Phys. Rev. Lett. \textbf{116}, 236401 (2016).

\bibitem{Kaganov1983}
M. I. Kaganov and A. A. Slutskin, Phys. Rep. \textbf{4}, 189 (1983).

\bibitem{Falicov1966}
L.M.\ Falicov and H.\ Stachowiak, Phys.\ Rev.\ \textbf{147}, 147 (1966).

\bibitem{Breitkreiz2018}
M.\ Breitkreiz, N.\ Bovenzi, and J.\ Tworzyd{\l}o , Phys.\ Rev.\ B \textbf{98}, 121403(R) (2018).

\bibitem{Stark1974}
R. W. Stark and C. B. Friedberg. Journal of Low Temperature Physics, \textbf{14} 111–146 (1974).

\bibitem{Stark1977}
R. W. Stark and R. Reifenberger, Journal of Low Temperature Physics, \textbf{26}, 819-826 (1977).


\end{thebibliography}
\end{document}